\documentclass[aps,pre,twocolumn,amsfonts,amssymb,floatfix,10pt,longbibliography]{revtex4-2}
\usepackage{hyperref}
\usepackage[T1]{fontenc}
\usepackage[utf8]{inputenc}
\usepackage{float}
\usepackage{graphicx}
\usepackage{amssymb}
\usepackage{amsmath}
\usepackage{enumerate}
\usepackage{relsize}

\usepackage{tikz}
\usetikzlibrary{decorations.pathmorphing,patterns}
\usetikzlibrary{arrows.meta}
\newcommand{\shift}{-6mm}
\newcommand{\Pa}{0.0, -0.5}
\newcommand{\Pb}{1.0, 0.0}
\newcommand{\Pc}{1.0, -0.5}
\newcommand{\Pd}{1.0,-1.0}
\newcommand{\Pe}{0.8, -1.3}
\newcommand{\Pf}{0.3, -1.5}
\newcommand{\focus}{\filldraw[black] (\Pa) circle (2pt);
		\draw[] (\Pa) circle (5pt);
		\draw  (-0.1,-0.5) node[anchor=east] {1};}
\newcommand{\focustwo}{\filldraw[black] (\Pa) circle (2pt);
		\draw[] (\Pa) circle (5pt);
		\draw  (-0.1,-0.5) node[anchor=east] {2};}
\newcommand{\focusnocircle}{\filldraw[black] (\Pa) circle (2pt);
		\draw  (-0.1,-0.5) node[anchor=east] {1};}
\newcommand{\PointB}[1]{\filldraw[black] (\Pb) circle (2pt) node[anchor=west] {#1};}
\newcommand{\PointC}[1]{\filldraw[black] (\Pc) circle (2pt) node[anchor=west] {#1};}
\newcommand{\PointD}[1]{\filldraw[black] (\Pd) circle (2pt) node[anchor=west] {#1};}
\newcommand{\PointE}[1]{\filldraw[black] (\Pe) circle (2pt) node[anchor=west] {#1};}
\newcommand{\PointF}[1]{\filldraw[black] (\Pf) circle (2pt) node[anchor=north] {#1};}

\newcommand{\Dash}[2]{\draw[dashed] (#1) to (#2);}
\newcommand{\DashC}[4]{\draw[dashed] (#1) to [out=#3, in=#4] (#2);}
\newcommand{\Sol}[2]{\draw[] (#1) to (#2);}
\newcommand{\SolC}[4]{\draw[] (#1) to [out=#3, in=#4] (#2);}
\newcommand{\Coil}[2]{\draw[decoration={coil}, decorate, segment length=4pt] (#1) to (#2);}
\newcommand{\CoilC}[4]{\draw[decoration={coil}, decorate, segment length=4pt] (#1) to [out=#3, in=#4](#2);}
\newcommand{\Arr}[2]{\draw[arrows={->[scale=1.7]}] (#1) to (#2);}
\newcommand{\ArrC}[4]{\draw[arrows={->[scale=1.7]}] (#1) to [out=#3, in=#4] (#2);}
\newcommand{\tz}[1]{\begin{tikzpicture}[baseline=\shift] #1 \end{tikzpicture}}

\usepackage{stackengine} 
\newcommand\kreis[1]{\stackMath\mathbin{\stackinset{c}{0ex}{c}{0ex}{#1}{\mathlarger{\mathlarger{\mathlarger{\mathlarger{\bigcirc}}}}}}}

\makeatletter

\newcommand*{\diff}{\mathop{\!\mathrm{d}\!}}

\newcount\hour \newcount\minute
\hour=\time \divide \hour by 60 \minute=\time \count99=\hour
\multiply \count99 by -60 \advance\minute by \count99 \hour=\time
\divide \hour by 60
\date{February 25, 2021} 

\makeatother

\begin{document}

\title{A Quantitative Kinetic Theory of Flocking with Three-Particle-Closure}

\author{Rüdiger Kürsten}
\affiliation{Institut für Physik, Universität Greifswald, Felix-Hausdorff-Str. 6, 17489 Greifswald, Germany}
\author{Thomas Ihle}
\affiliation{Institut für Physik, Universität Greifswald, Felix-Hausdorff-Str. 6, 17489 Greifswald, Germany}

\begin{abstract}
We consider aligning self-propelled particles in two dimensions. 
Their motion is given by generalized Langevin equations and includes non-additive $N$-particle interactions.
The qualitative behavior is as for the famous Vicsek model.
We develop a kinetic theory of flocking beyond mean field.
In particular, we self-consistently take into account the full pair correlation function.
We find excellent quantitative agreement of the pair correlations with direct agent-based simulations within the disordered regime.
Furthermore we use a closure relation to incorporate spatial correlations of three particles.
In that way we achieve good quantitative agreement of the onset of flocking with direct simulations.
Compared to mean field theory, the flocking transition is shifted significantly towards lower noise because directional correlations favor disorder.
We compare our theory with a recently developed Landau- kinetic theory.
\end{abstract}
\maketitle

\section{Introduction}

Entities equipped with a propulsion mechanism, that is active matter, transfer free energy into directed motion.
Such entities exist from micro or even nano length scales up to macroscopic size.
Examples are man-made micro swimmers, bacteria, insects, fish, mammals and robots.

Large groups of interacting active particles can exhibit complex emergent collective phenomena that differ fundamentally from their equilibrium counter parts.
Some prominent examples are flocking \cite{VCBCS95, TT95}, motility induced phase separation \cite{FM12, BBKLBS13}, and bacterial turbulence \cite{Wolgemuth08}.
An overview on the rapidly developing field of active matter and its application can be found e.g. in the following reviews \cite{TTR05, Ramaswamy10, VZ12, RBELS12, Klotsa19, DL19, Chate20, SWWGR20, BGHP20}.

The Vicsek model \cite{VCBCS95} is historically one of the first and computationally one of the simplest active models that exhibits a flocking transition. Therefore it is considered as one of the prototype models of active matter.
In this model, point particles move in two dimensions at constant speed $v_0$ in individual directions given by the polar angles $\phi_i$, $i \in \{1, \dots, N\}$.
At points in time $\tau, 2\tau, 3\tau, \dots$ all particle directions $\phi_i$ are changed due to interactions in the following way:
Each particle takes the direction of the average velocity of all particles within distance $R$ (including the particle itself), disturbed by a noise term.
Thus, the interactions favor a local alignment of the particle velocities.

For large systems and periodic boundary conditions the Vicsek model is known in four phases \cite{VCBCS95, GC04, CGGR08, SCT15, Chate20, KI20}.
For small noise or large particle densities, on average all particles move in a similar direction (i) \cite{VCBCS95}.

Hence we call the system polarly ordered.
Furthermore, particles tend to cluster together locally.
However, the clusters are distributed equally over space and thus the particle distribution is homogeneous on larger length scales.
By increasing the noise strength or decreasing the particle density the system arranges in a 'cross sea' phase (ii) \cite{KI20}, where high density regions that look like crossing wave fronts are formed.
This pattern moves through the system and the low density regions have almost no polar order.
For even larger noise strength or smaller densities the system arranges in parallel non-intersecting traveling high density bands (iii) \cite{GC04}. 
Again, there is almost no polar order in the low density regions between the high density bands.
In phases (ii) and (iii) there is still an average polar order. 
The main contributions to the polar order come from the high density regions.
For very large noise strength or very small particle densities there is no polar order (iv).
That is, for large systems there is no motion of the center of mass and the particles are distributed homogeneously.

There have been qualitative descriptions of some of the aforementioned phases by field- or kinetic theories \cite{TT95, TT98, Ihle11, Ihle13, Ihle16}.
Those theories either do not reach quantitative agreement with direct simulations or only for very special parameters.
In addition, most theories rely on the mean field assumption.
There are also kinetic theories of active systems that consider weak pair correlations to some extent, see e.g. \cite{CI15, SNNMM17, SNSMM20, Patelli20}.
However, for the Vicsek model, recently it has been shown quantitatively, that correlations of two and more particles are important even in large parts of the disordered phase (iv) \cite{KSZI20}.

The aim of this paper is to provide a more quantitative theoretical description of the model.
For technical reasons we are not studying the Vicsek model itself but a similar model that is believed to behave qualitatively equivalent to the Vicsek model.
Major problems in kinetic theories of the standard Vicsek model are difficulties related to the finite time step as well as the presence of multi-particle collision integrals that are not analytically solvable, see e.g. \cite{Ihle11}.
The latter was circumvented in \cite{CI15} by using binary interactions with randomly selected interaction partners.

In this paper we develop a ring-kinetic theory. 
That means, we consider the dynamical equations for the one-particle distribution and for the two-particle correlation function explicitly.
Higher order correlations are neglected in the first step.
However, the effects of the pair correlations on the one-particle distribution as well as on the pair correlation function itself, are fully taken into account.
This concept has been applied in various fields, see e.g. \cite{EC81, Leutheusser83, BED95, NEB98}.
In this paper, we set up the ring-kinetic equations for $N$-particle interactions, a significant difference to most previous applications.
That means the forces do not depend on the state of only two particles but on the positions of all particles.

Within the disordered parameter regime, we find excellent quantitative agreement between the ring-kinetic theory and agent-based simulations.
For a moderate particle density, we find the threshold noise for the onset of flocking within the ring-kinetic theory to be very close to the value measured in agent-based simulations.
In particular, the results are much more precise than the predictions of mean field theory.

For larger particle densities, at the onset of flocking, higher order correlations are more important.
We extend the ring-kinetic theory and incorporate also spatial three-particle correlations via a closure ansatz.
In that way, the flocking transition can be described also for larger densities, and the agreement between theory and simulations is further improved at moderate densities, where we reach quantitative agreement within the considered resolution.
In addition, for higher densities, our theory is a significant improvement over mean field.

The paper is organized as follows.
In Sec.~\ref{sec:model} we define and discuss the Vicsek-like model studied here.
In Sec.~\ref{sec:kinetic_theory} we develop a homogeneous mean field theory and calculate the critical noise strength of the flocking transition.
In Sec.~\ref{sec:correlations} we introduce the notion of many particle correlations.
In Sec.~\ref{sec:ring_kinetic_theory} we develop the full ring-kinetic theory, that is the dynamical equations for the one particle distribution and the pair correlation function neglecting higher order correlations.
In Sec.~\ref{sec:numerics} we present extensive quantitative comparisons between the solutions of the ring-kinetic equations and direct agent-based simulations.
We identify the parameter region where the ring-kinetic theory is applicable.
In Sec.~\ref{sec:closure_relation} we employ a closure relation to take into account spatial three particle correlations.
In that way we can significantly enlarge the applicability domain of the kinetic theory.
In Sec.~\ref{sec:discussion} we discuss our results and give an outlook to possible extensions of the method.
In Appendix~\ref{sec:fourier} we give all relevant equations in Fourier space that have been used to evaluate the kinetic equations numerically.
In Appendix~\ref{sec:landau} we compare the results of the kinetic theory presented in this work with the simplified Landau kinetic theory of Patelli \cite{Patelli20}.

\section{Model\label{sec:model}}

We consider a Vicsek-like model in continuous time that was investigated as here, or in a similar form in many studies, see e.g. \cite{PDB08, PSB10, FMMT12, LL17}. The model is given by
\begin{align}
	\dot{x}_{i} &= v_0 \cos(\phi_{i}) 
	\notag
	\\
	\dot{y}_{i} &= v_0 \sin(\phi_{i}) 
	\notag
	\\
	\dot{\phi}_{i} &= w(|\Omega(i)|) \sum_{j\in \Omega(i)} \sin(\phi_j-\phi_i) + \sigma \xi_{i}, \quad i=1, \dots, N,
	\label{eq:model}
\end{align}
where $\mathbf{r}_i(t)=(x_i(t), y_i(t))$ denote the particles positions and  $\phi_i$ the directions of the particles velocities. 
The set
\begin{align}
	\Omega_i:=\{j \in \{1, \dots, N\}: |\mathbf{r}_j-\mathbf{r}_i|\le R\}
\end{align}
contains all particles that are within distance $R$ to particle $i$.
The $\xi_{i}(t)$ are independent Gaussian white noise terms satisfying 
\begin{align}
	\langle \xi_{i}(t) \xi_j(s) \rangle = \delta_{ij}\delta(t-s).
	\label{eq:whitenoise}
\end{align}
The noise strength is given by $\sigma$, and $w(n)$ is an interaction weight function that depends on the number of neighbors of particle $i$ (particles that are within distance $R$ including particle $i$ itself).

We consider the two-dimensional motion of $N$ particles that move at constant speed $v_0$ in individual directions $\phi_i$.
The directions of neighboring particles tend to align, however, they are disturbed by noise.
In the following, we discuss the cases $w(n)=1=const.$ and $w(n)=1/n$. 

From a technical point of view, the case $w(n)=1=const.$ is the most desirable
since in this case, the model includes only pair interactions.
This has the advantage that, as in the regular BBGKY-hierarchy, three-particle correlations can be produced only from previously existing pair correlations, four-particle correlations can only be produced from three-particle correlations and so on.
In contrast, if $w$ is a function of $n$ as e.g. $w(n)=1/n$ the interactions are in fact $N$-particle interactions.
That means, all orders of correlations are produced immediately even if the model evolves from uncorrelated initial conditions.

\begin{figure}
	\includegraphics[width=0.46\columnwidth]{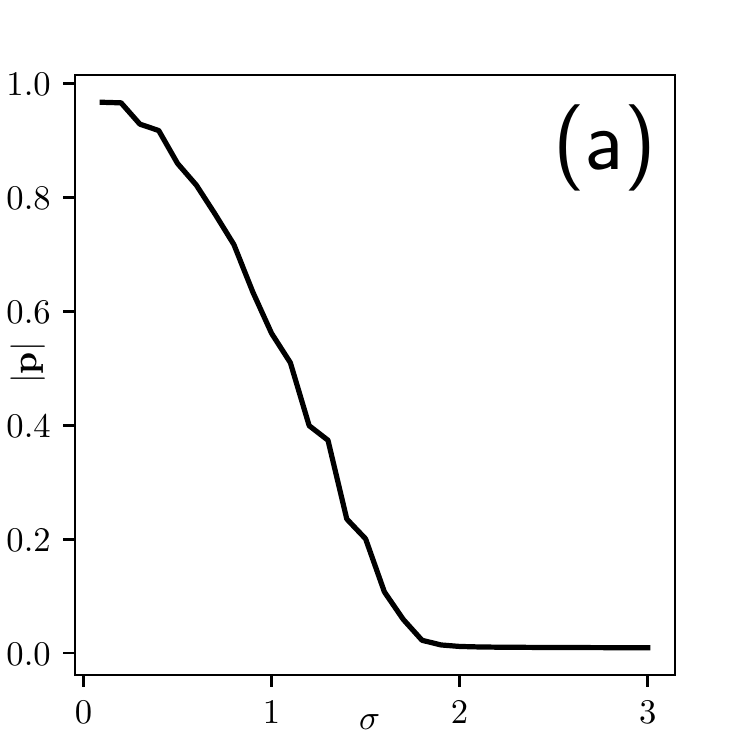}
	\includegraphics[width=0.46\columnwidth]{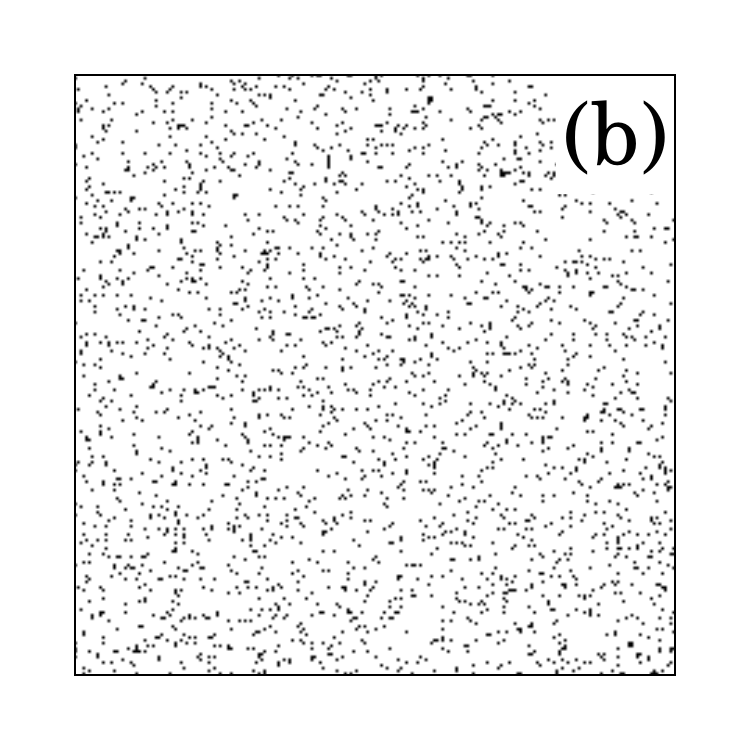}\\
	\includegraphics[width=0.46\columnwidth]{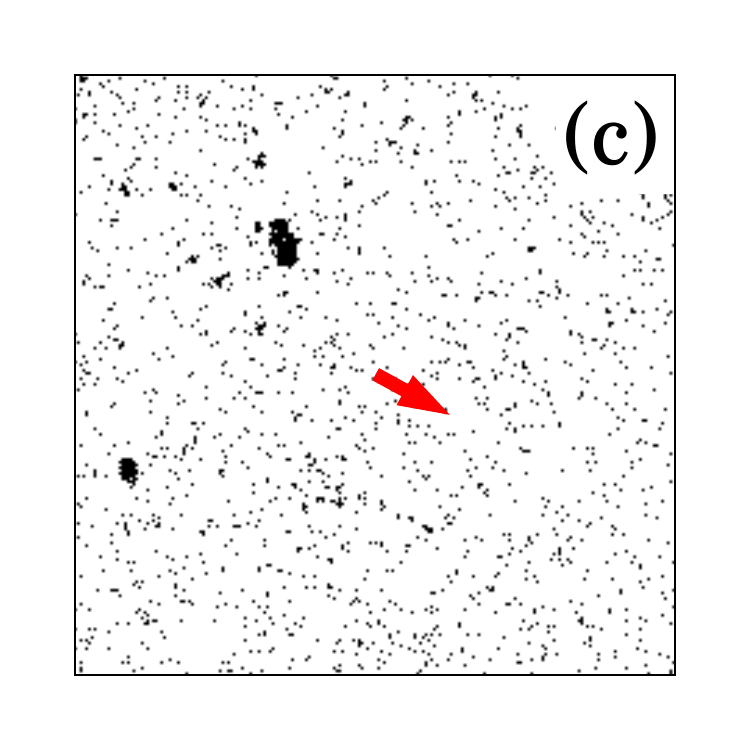}
	\includegraphics[width=0.46\columnwidth]{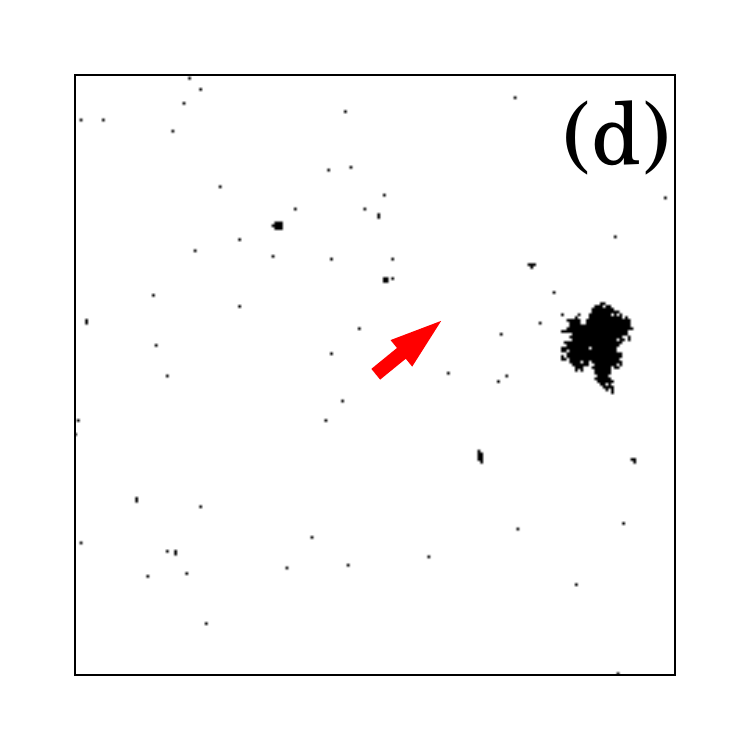}
	\caption{Numerical results for the system \eqref{eq:model} with $w(n)=1=const.$, simulated with an Euler-Maruyama-scheme \protect{\cite{KP92}} with $\Delta t=0.01$. Parameters are $N=10^4$, $L\approx 177.25$, $v=R=1$, thus $C_1:=\rho_0 \pi R^2=1$.  (a) Polar order parameter as a function of noise strength. (b-d) Snapshots after a thermalization time $T=1000$ in the disordered phase (b) $\sigma=2.2$, and in the polarly ordered phase (c) $\sigma=1.5$ and (d) $\sigma=0.3$.\label{fig:nodivision}}
\end{figure}

Intuitively, one might think that such tiny details of the model are not that important and might lead to qualitatively equivalent results.
The average number of neighbors is a constant anyway and one might hope that the fluctuations of the number of neighbors do not play a major role.
Surprisingly, that is not the case and the two models differ qualitatively, see also \cite{CSP21} for a detailed analysis.
For $w(n)=1=const.$ one still finds a homogeneous phase at large noise and a polarly ordered phase at smaller noise for systems of finite size.
However, in the case of polar order, particles do not arrange in high density bands or cross sea patterns but they form high density clusters that contain almost all particles, see Fig.~\ref{fig:nodivision}.

It is not even clear if there is a disordered phase at all in the thermodynamic limit.
An alternative hypothesis is as follows. For every finite but possibly large noise strength one finds polar ordered clusters of very high densities for large enough systems. 
Due to the additive nature of the alignment interactions, such a cluster can remain stable even for large noise strength if the density is large enough.
It could be that such a clustered state is the steady state for large systems at any noise strength.
However, sophisticated investigations are necessary to answer this question.
Differences between additive and non-additive interactions have been noticed already in \cite{Stroteich19}, they have been studied in more detail in a very recent work \cite{CSP21}.
Although the model with additive interactions is very interesting as well, we are mainly interested in the study of models behaving qualitatively like the Vicsek model.
Hence we focus on the case 
\begin{align}
	w(n)=1/n
	\label{eq:weight_function}
\end{align}
in this paper.

\section{Kinetic Theory\label{sec:kinetic_theory}}

Instead of investigating the Langevin equations, Eq.~\eqref{eq:model}, we can equivalently study the associated $N$-particle Fokker-Planck equation \cite{Risken89}
\begin{align}
	& \partial_t P_N(1, 2, \dots, N, t) = 
	\notag
	\\
	&- \sum_{i=1}^{N} \partial_{\phi_{i}}[ \sum_{j=1}^{N} w(|\Omega(i)|) \theta_{ij} 
	\notag
	\\
	&\times \sin(\phi_j-\phi_i) P_N(1, 2, \dots, N, t) ] 
	\notag
	\\
	&-\sum_{i=1}^{N} v \cos(\phi_i) \partial_{x_i} P_N(1, 2, \dots, N, t) 
	\notag
	\\
	&-\sum_{i=1}^{N} v \sin(\phi_i) \partial_{y_i} P_N(1, 2, \dots, N, t)
	\notag
	\\
	&+ \sum_{i=1}^{N}\frac{\sigma^2}{2} \partial_{\phi_i}^{2} P_N(1, 2, \dots, N, t),
	\label{eq:FPfull}
	\\
	&\theta_{ij}:= \theta(R- \sqrt{(x_j-x_i)^2+(y_j-y_i)^2}),
	\label{eq:thetaij}
\end{align}
where we used the abbreviation of writing just $1$ instead of $x_1, y_1, \phi_1$ and similar for $2, 3, \dots$. The probability density of finding the system in a phase space volume around the coordinates given by $1, 2, \dots, N$ at time $t$ is denoted by $P_N(1, 2, \dots, N, t)$.
This equation is exact.

We assume that the probability density is initially and hence at all times symmetric with respect to the permutation of coordinates $i \leftrightarrow j$.
We denote the marginalized probability density by
\begin{align}
	P_{k}(1, 2, \dots, k, t):= \int_{}^{} P_N(1, 2, \dots, N, t) \diff \, (k+1) \dots \diff N,
	\label{eq:margin}
\end{align}
where $\diff l$ denotes $\diff x_l \diff y_l \diff \phi_l$ and integration is performed over the interval $[0,L)$ for $x_l$ and $y_l$ and over the interval $[0, 2\pi)$ for $\phi_l$.

Using the Fokker-Planck equation \eqref{eq:FPfull} and the symmetry of permutations of coordinates we obtain the time evolution of the marginalized density
\begin{align}
	& \partial_t P_1(1, t) = 
	\notag
	\\
	&- \partial_{\phi_{1}}[ (N-1)   \int_{}^{} w(|\Omega(1)|) \diff 2 \dots \diff N  \theta_{12}
	\notag
	\\
	&\qquad \times \sin(\phi_{2}-\phi_1) P_{N}(1, 2, \dots, N, t) ] 
	\notag
	\\
	&- v \cos(\phi_1) \partial_{x_1} P_1(1, t) 
	- v \sin(\phi_1) \partial_{y_1} P_1(1, t)
	\notag
	\\
	&+ \frac{\sigma^2}{2} \partial_{\phi_1}^{2} P_1(1, t).
	\label{eq:FPmargin}
\end{align}
We observe that the time evolution of the one particle probability depends on the complete $N$ particle probability
due to the fact that the number of neighbors $|\Omega(1)|$ of the first particle depends on the positions of all the other particles.

For spatially homogeneous solutions, $P_1$ depends only on $\phi$ and $t$, hence we can define
\begin{align}
	p_{\phi}(\phi_1, t) := L^2 P_{1}(1, t)
	\label{eq:homogeneousP}
\end{align}
such that Eq.~\eqref{eq:FPmargin} simplifies to 
\begin{align}
	& \partial_t p_\phi(\phi_1, t) = 
	\notag
	\\
	&- \partial_{\phi_{1}}[ (N-1)   \int_{}^{} w(|\Omega(1)|) \diff 1 \dots \diff N \, \theta_{12}  
	\notag
	\\
	&\qquad \times \sin(\phi_{2}-\phi_1) P_{N}(1, 2, \dots, N, t) ] 
	\notag
	\\
	&+ \frac{\sigma^2}{2} \partial_{\phi_1}^{2} p_\phi(\phi_1, t),
	\label{eq:FPmargin2}
\end{align}

Considering the noise strength $\sigma$ as control parameter, in the thermodynamic limit $N\rightarrow \infty$, $\frac{N}{L^2}=const.$, the system exhibits a transition from disorder to polar order.
In the disordered phase, all particles directions are distributed randomly. 
In the polar ordered phase there is a global preferred direction of motion.

In the next section we calculate the threshold noise strength where the transition takes place, within the mean field theory.

\subsection{Mean Field Theory}

Solving Eq.~\eqref{eq:FPmargin2} exactly is very difficult or even impossible.
Therefore we are looking for the stationary solution under the mean field assumption 
\begin{align}
	P_{N}(1, 2, \dots, N)= P_1(1)P_1(2)\dots P_1(N),
	\label{eq:meanfieldassumption}
\end{align}
which simplifies with Eq.~\eqref{eq:homogeneousP} to
\begin{align}
	P_N(1, 2, \dots, N)
	&=\frac{1}{L^{2N}}p_{\phi}(\phi_1) p_{\phi}(\phi_2)\dots p_{\phi}(\phi_N)
	\label{eq:meanfieldassumption3}
\end{align}
In order to simplify Eq.~\eqref{eq:FPmargin2} it is reasonable to split the spatial integration domain into pieces such that the number of neighbors of particle one is constant on each piece.
Inserting this assumption into Eq.~\eqref{eq:FPmargin2} we obtain
\begin{align}
	& \partial_t p_\phi(\phi_1, t) = 
	\notag
	\\
	&- \partial_{\phi_{1}}\bigg[ \frac{N-1}{L^{2(N-1)}}   \sum_{n=1}^{N-1}\int_{}^{} w(n+1) \diff 2 \diff \mathbf{r}_3 \dots \diff \mathbf{r}_N  \theta_{12} \dots \theta_{1(n+1)}
	\notag
	\\
	&\times (1-\theta_{1(n+2)})\dots (1-\theta_{1N})\sin(\phi_{2}-\phi_1)
	\notag
	\\
	&\qquad \times  p_{\phi}(\phi_1, t) p_{\phi}(\phi_2, t) + permutations \bigg] 
	\notag
	\\
	&+ \frac{\sigma^2}{2} \partial_{\phi_1}^{2} p_\phi(\phi_1, t),
	\label{eq:FPmargin3}
\end{align}
where $\mathbf{r}_i=(x_i, y_i)$. In the above integral $\mathbf{r}_2, \dots, \mathbf{r}_{n+1}$ are integrated over the inside of the disk with radius $R$ around particle one and $\mathbf{r}_{n+2} , \dots, \mathbf{r}_N$ are integrated over the outside of the disk around particle one. 
That means $n$ is the number of neighbors of particle one.
By permutations we mean all other choices of $n-1$ particles from $\{3, \dots, N\}$ to be integrated over the inside.
All this permutations give the same contribution, hence we can take care of them by a combinatorial factor to arrive at
\begin{align}
	& \partial_t p_\phi(\phi_1, t) = 
	\notag
	\\
	&- \partial_{\phi_{1}}\bigg[ \frac{N-1}{L^{2(N-1)}}  \sum_{n=1}^{N-1}\binom{N-2}{n-1} \int_{}^{} w(n+1) \diff 2 \diff \mathbf{r}_3 \dots \diff \mathbf{r}_N  
	\notag
	\\
	&\times \theta_{12} \dots \theta_{1(n+1)}(1-\theta_{1(n+2)})\dots (1-\theta_{1N})\sin(\phi_{2}-\phi_1)
	\notag
	\\
	&\qquad \times  p_{\phi}(\phi_1, t) p_{\phi}(\phi_2, t) \bigg] 
	+ \frac{\sigma^2}{2} \partial_{\phi_1}^{2} p_\phi(\phi_1, t).
	\label{eq:FPmargin4}
\end{align}
Here, in the two dimensional case, the $C_1$ coefficient is calculated as
\begin{align}
	C_1:= \frac{N}{L^2}\int_{}^{} \theta_{12} \diff \mathbf{r}_2,
	\label{eq:densityM}
\end{align}
that is the expectation value of the number of particles within a circle of radius $R$.
Since we are interested in the thermodynamic limit $N \rightarrow \infty$ it suffices to approximate
\begin{align}
	(N-1) \binom{N-2}{n-1} \approx \frac{N^n}{(n-1)!}.
	\label{eq:binomapprox}
\end{align}
Inserting Eqs.~\eqref{eq:densityM} and \eqref{eq:binomapprox} into Eq.~\eqref{eq:FPmargin4} we obtain
\begin{align}
	& \partial_t p_\phi(\phi_1, t) = 
	- \sum_{n=1}^{N-1} \frac{C_1^n}{(n-1)!} \big(1- \frac{C_1}{N}\big)^{N-1-n} w(n+1)
	\notag
	\\
	&\times\partial_{\phi_{1}}\bigg[ \int_{}^{}  \diff \phi_2   \sin(\phi_{2}-\phi_1) p_{\phi}(\phi_1, t) p_{\phi}(\phi_2, t) \bigg] 
	\notag
	\\
	&+ \frac{\sigma^2}{2} \partial_{\phi_1}^{2} p_\phi(\phi_1, t).
	\label{eq:FPmargin5}
\end{align}
Taking the limit $N \rightarrow \infty$ and substituting $k=n-1$ we arrive at
\begin{align}
	& \partial_t p_\phi(\phi_1, t) = 
	- C_1 \sum_{k=0}^{\infty} \frac{C_1^k}{k!} \exp(-C_1) w(k+2)
	\notag
	\\
	&\times\partial_{\phi_{1}}\bigg[ \int_{}^{}  \diff \phi_2   \sin(\phi_{2}-\phi_1) p_{\phi}(\phi_1, t) p_{\phi}(\phi_2, t) \bigg] 
	\notag
	\\
	&+ \frac{\sigma^2}{2} \partial_{\phi_1}^{2} p_\phi(\phi_1, t).
	\label{eq:FPmargin6}
\end{align}
One way of treating Eq.~\eqref{eq:FPmargin6} is to study its full Fourier transform.
Since here, we are only interested in the critical noise strength it suffices to consider only the zeroth and the first order in the Fourier transform.
Rotating the coordinate system in an appropriate way we can use the ansatz
\begin{align}
	p_{\phi}(\phi) = \frac{1}{2\pi} + \varepsilon \cos(\phi).
	\label{eq:ansatzorder}
\end{align}
The isotropic distribution corresponding to $\varepsilon=0$ is always a solution of Eq.~\eqref{eq:FPmargin6}.
At the critical noise strength it changes stability.

Inserting the ansatz \eqref{eq:ansatzorder} into Eq.~\eqref{eq:FPmargin6}, multiplying by $\cos(\phi_1)/\pi$ integrating over $\phi_1$ and neglecting terms of order $\varepsilon^2$ we obtain by partial integration
\begin{align}
	& \partial_t \varepsilon = 
	- \frac{1}{\pi} C_1 \sum_{k=0}^{\infty} \frac{C_1^k}{k!} \exp(-C_1) w(k+2)
	\notag
	\\
	&\times \bigg\{ \int_{0}^{2\pi} \diff \phi_1 \diff \phi_2 \sin(\phi_1)  \sin(\phi_{2}-\phi_1) 
	\notag
	\\
	&\times \bigg[\frac{1}{4\pi^2} + \frac{\varepsilon \cos(\phi_1)}{2\pi} + \frac{\varepsilon \cos(\phi_2)}{2\pi}   \bigg] \bigg\} 
	- \frac{\sigma^2}{2} \varepsilon
	\notag
	\\
	&= \bigg[ \frac{C_1}{2}  \sum_{k=0}^{\infty} \frac{C_1^k}{k!} \exp(-C_1) w(k+2) -\frac{\sigma^2}{2} \bigg] \varepsilon.
	\label{eq:timeevolepsilon}
\end{align}
Hence the mean field critical noise strength is
\begin{align}
	\sigma_c= \sqrt{C_1  \sum_{k=0}^{\infty} \frac{C_1^k}{k!} \exp(-C_1) w(k+2)}.
	\label{eq:sigmacmf}
\end{align}
In case of the constant weight function $w(n)=1$ this reduces to
\begin{align}
	\sigma_c= \sqrt{C_1}.
	\label{eq:sigmacmf1}
\end{align}
For a Poisson distribution we have
\begin{align}
	&\bigg\langle \frac{1}{k+1}\bigg\rangle = \sum_{k=0}^{\infty} \frac{1}{k+1} \frac{C_1^k}{k!}\exp(-C_1)
	\notag
	\\
	&=\frac{1}{C_1} \sum_{k=0}^{\infty} \frac{C_1^{k+1}}{(k+1)!}\exp(-C_1)
	\notag
	\\
	&=\frac{1}{C_1} \sum_{\tilde{k}=1}^{\infty} \frac{C_1^{\tilde{k}}}{\tilde{k}!}\exp(-C_1)
	\notag
	\\
	&=\frac{1}{C_1} \bigg[ \sum_{\tilde{k}=0}^{\infty} \frac{C_1^{\tilde{k}}}{\tilde{k}!}\exp(-C_1) -\exp(-C_1) \bigg]
	\notag
	\\
	&=\frac{1}{C_1} [ 1 -\exp(-C_1) ]
	\label{eq:poissonexpr1}
\end{align}
and similar
\begin{align}
	&\bigg\langle  \frac{1}{k+2} \bigg\rangle = \bigg\langle \frac{1}{k+1}\bigg\rangle - \bigg\langle \frac{1}{(k+1)(k+2)} \bigg\rangle
	\notag
	\\
	&=\bigg\langle \frac{1}{k+1}\bigg\rangle -\frac{1}{C_1^2} \sum_{\tilde{k}=2}^{\infty} \frac{C_1^{\tilde{k}}}{\tilde{k}!} \exp(-C_1)
	\notag
	\\
	&=\bigg\langle \frac{1}{k+1}\bigg\rangle - \frac{1}{C_1^2}[1-(1+C_1)\exp(-C_1)].
	\label{eq:poissonexpr2}
\end{align}
Hence, for the weight function $w(n)=1/n$ we obtain the critical noise strength using Eqs.~\eqref{eq:sigmacmf}, \eqref{eq:poissonexpr1} and \eqref{eq:poissonexpr2}
\begin{align}
	\sigma_c &= \sqrt{\{ 1-\exp(-C_1) - \frac{1}{C_1} [1-(1+C_1)\exp(-C_1)]\}}
	\notag
	\\
	&=\sqrt{ 1 - \frac{1}{C_1} [1-\exp(-C_1)]}.
	\label{eq:sigmacmf2}
\end{align}

\section{Correlations\label{sec:correlations}}

In general, the mean field assumption \eqref{eq:meanfieldassumption} is not valid but must be corrected by terms containing correlations of various order.
The correlation functions $G_k$ are defined recursively by \cite{Ursell27, MM41, EC81}
\begin{align}
	&G_1\equiv P_1,\\
	&G_k(1, \cdots, k):= P_k(1, \cdots, k) 
	\label{eq:correlationfunction}
	\\
	&- \sum_{\sigma}\sum_{l=1}^{k-1} \frac{1}{(l-1)!} \frac{1}{(k-l)!}G_{l}(1, {\sigma(2)}, \cdots, {\sigma(l)}) \notag
	\\
	&\times P_{k-l}({\sigma(l+1)}, \cdots, {\sigma(k)}),
	\notag
\end{align}
where $\sum_\sigma$ denotes the sum over all permutations of the elements $\{2, \cdots, k\}$.
We can rewrite Eq.~\eqref{eq:correlationfunction} as $P_l(\dots) = G_l(\dots) + \dots$ and insert it recursively for all $P_l$ of order $l<k$ into Eq.~\eqref{eq:correlationfunction} and eventually replace $P_1$ by $G_1$. Performing such an expansion, only $P_k$ and $G$-functions remain on the right hand side of Eq.~\eqref{eq:correlationfunction}.
It follows inductively from Eq.~\eqref{eq:correlationfunction} that the indexes of all correlation functions on the right hand side are ordered, that is for each term $G_l({i_1}, \dots, {i_l})$ appearing in the expansion of Eq.~\eqref{eq:correlationfunction} it holds $i_1<i_2<\dots<i_l$.
Thus, instead of Eq.~\eqref{eq:correlationfunction} we might alternatively write
\begin{align}
	&G_k(1, \dots, k)=P_k(1, \dots, k) - \sum \{ \text{over all possible}
	\notag\\
	&\text{products of $G$-functions such that each of the arguments}
	\notag\\
	&\text{$1, \dots, k$ appears exactly once and for each $G$-function}
	\notag
	\\
	&\text{the arguments are ordered.}\}
	\label{eq:alternativedefinitioncorrelationfunction}
\end{align}
For example, the two-, three- and four- particle correlation functions are given explicitly as
\begin{align}
	&G_2(1, 2)=P_2(1, 2)-P_1(1)P_1(2)
		\label{eq:g2}
	\\
	&G_3(1, 2, 3)=P_3(1, 2, 3)-P_1(1)P_1(2)P_1(3)
	\notag
	\\
	&-P_1(1)G_2(2, 3)-P_1(2)G_2(1, 3)-P_1(3)G_2(1, 2),
		\label{eq:g3}
		\\
	&G_4(1, 2, 3, 4)=P_4(1, 2, 3, 4)-P_1(1)P_1(2)P_1(3)P_1(4)
	\nonumber
	\\
	&-G_2(1,2)P_1(3)P_1(4)-G_2(1,3)P_1(2)P_1(4)
	\nonumber
	\\
	&-G_2(1,4)P_1(2)P_1(3)-G_2(2,3)P_1(1)P_1(4)
	\nonumber
	\\
	&-G_2(2,4)P_1(1)P_1(3)-G_2(3,4)P_1(1)P_1(2)
	\nonumber
	\\
	&-G_2(1,2)G_2(3, 4)-G_2(1,3)G_2(2, 4)-G_2(1,4)G_2(2, 3)
	\nonumber
	\\
	&-G_3(1, 2, 3)P_1(4)-G_3(1, 2, 4)P_1(3)
	\nonumber
	\\
	&-G_3(1, 3, 4)P_1(2)-G_3(2, 3, 4)P_1(1).
		\label{eq:g4}
\end{align}
We want to mention some important properties of the correlation functions $G_k$.
Since we assume that the $N$-particle probability distribution is symmetric with respect to particle exchanges, also the correlation functions follow the same symmetry.
Furthermore, for $k\ge 2$ it holds
\begin{align}
	\int_{}^{} G_k(1, \dots, k) \diff k =0,
	\label{eq:correlationintegral}
\end{align}
where the integration is performed over the whole space and all possible orientations of particle $k$.
This property follows by induction over $k$ from Eq.~\eqref{eq:correlationfunction}.
It shows that $G_k$ contains only information about $k$-particle correlations and not about lower order correlations.
As a simple consequence it follows that
\begin{align}
	\int_{X}^{} G_k(1, \dots, k) \diff k = -\int_{X^C}^{} G_k(1, \dots, k) \diff k ,
	\label{eq:correlationintegralsubset}
\end{align}
where $X$ is some subset of the set of all possible configurations of particle $k$ and $X^C$ is the complement of $X$.

\section{Ring-Kinetic Theory\label{sec:ring_kinetic_theory}}

The mean field assumption \eqref{eq:meanfieldassumption} is equivalent to vanishing correlation functions $G_i\equiv 0$ for $i\ge 2$.
In this paper we go one step further and take two particle correlations into account.
However, we still assume that higher order correlations can be neglected.
That is, we consider
\begin{align}
	P_1(1, t)&,
	\notag
	\\
	G_2(1, 2, t)&:= P_2(1, 2, t) - P_1(1,t)P_1(2, t),
	\notag
	\\
	G_i(1, \cdots, i, t) &\equiv 0 \,\,\,\,\text{ for }i\ge 3.
	\label{eq:ringkinetic}
\end{align}
Similar assumptions have been made in Ref. \cite{CI15} but also in a very recent kinetic Landau theory \cite{Patelli20} on the model with additive interactions, that is with $w(n)\equiv 1$.
Therein, it was assumed additionally that $G_2$ is small and furthermore that also noise and coupling are small.
It should be mentioned that according to \cite{KSZI20} and to the results presented in this work, it follows that in the vicinity of the flocking transition and even in a considerable part of the disordered phase, the assumption of negligible $G_3$ and small $G_2$ is not justifiable.
However, far in the disordered regime it is a reasonable approximation.
In Appendix \ref{sec:landau} we present a detailed comparison of the Landau theory of \cite{Patelli20} applied to the model of non-additive interactions with our full ring-kinetic theory.

In this paper we impose no restrictions on $G_2$, that means we allow large pair correlations.
As we assume spatial homogeneity, the one-particle probability density $P_1$ is independent on the position $(x_1, y_1)$ and it suffices to consider the angular dependence $p_\phi$ given by Eq.~\eqref{eq:homogeneousP}.
Similarly, the two-particle probability density $P_{2}$ depends only on the angles $\phi_1, \phi_2$ and on the distance between the spatial coordinates $\Delta_x:=x_2-x_1, \Delta_y:=y_2-y_1$, $\Delta=(\Delta_x, \Delta_y)$.
Hence it is reasonable to introduce the reduced probability density
\begin{align}
	&p_2(\phi_1, \phi_2, \Delta) := L^2 \int_{}^{} P_2(\phi_1, \phi_2, x_1, x_2, y_1, y_2)
	\notag
	\\
	&\times\delta(\Delta_x -(x_2-x_1)) \delta(\Delta_y-(y_2-y_1)) \diff x_1 \diff x_2 \diff y_1 \diff y_2
	\notag
	\\
	&=L^4 P_2(\phi_1, \phi_2, x_1, x_2=x_1+\Delta_x, y_1, y_2=y_1+\Delta_y)
	\label{eq:reduced_ptwo}
\end{align}
and correlation functions
\begin{align}
	&g_2(\phi_1, \phi_2, \Delta) := L^2 \int_{}^{} G_2(\phi_1, \phi_2, x_1, x_2, y_1, y_2)
	\notag
	\\
	&\times\delta(\Delta_x -(x_2-x_1)) \delta(\Delta_y-(y_2-y_1)) \diff x_1 \diff x_2 \diff y_1 \diff y_2
	\notag
	\\
	&=L^4 G_2(\phi_1, \phi_2, x_1, x_2=x_1+\Delta_x, y_1, y_2=y_1+\Delta_y).
	\label{eq:reduced_gtwo}
\end{align}
We obtain the time evolution equations of the reduced one- and two-particle probability density functions, that are the first two equations of the BBGKY-hierarchy, from the Fokker-Planck equation \eqref{eq:FPfull} using the marginalization \eqref{eq:margin}, plugging in the cluster expansion \eqref{eq:correlationfunction} and using the ring-kinetic ansatz \eqref{eq:ringkinetic} as well as properties \eqref{eq:correlationintegral} and \eqref{eq:correlationintegralsubset} of the correlation functions.
We use the following diagramatic representation of the appearing collision integrals
\begin{align} 
	\tz{\focusnocircle \PointC{2} \Arr{\Pa}{\Pc}}&=\sin(\phi_2-\phi_1)\theta(R-|\mathbf{r}_2-\mathbf{r}_1|),\\
	\tz{\focusnocircle \PointC{2} \Sol{\Pa}{\Pc}}&=\theta(R-|\mathbf{r}_2-\mathbf{r}_1|),\\
	\tz{\focusnocircle \PointC{2} \Dash{\Pa}{\Pc}}&=1-\theta(R-|\mathbf{r}_2-\mathbf{r}_1|),\\
	\tz{\focusnocircle \PointC{2} \Coil{\Pa}{\Pc}}&=g_2(1,2),\\
	\tz{\PointC{k}}&=p(\phi_k), \text{if }k\text{ is not connected}\\
	&\phantom{=p_{\phi}(\phi_k), } \text{  via a coiled line,} \notag \\
	\tz{\focus} &= \partial_{\phi_1}.
\end{align}
All terms representing a symbol in a diagram are meant to be multiplied and integrated over all degrees of freedom that do not occur on the left hand side of the equation in which the diagram appears.
As an example, we give the meaning of the following integral
\begin{align}
	\tz{\focus \PointC{2} \ArrC{\Pa}{\Pc}{30}{150} \CoilC{\Pa}{\Pc}{330}{210}}= \int &\partial_{\phi_1} g_2(\phi_1, \phi_2, \Delta)\sin(\phi_2-\phi_1)
	\notag \\
	&\times \theta(R-|\Delta|) \diff \phi_2 \diff \Delta.
	\label{eq:bsp}
\end{align}
For the one particle angular distribution we find within our diagramatic notation 
\begin{align}
	&\partial_t p(\phi_1)= -\rho w_2 \bigg[
	\tz{\focus \PointC{3} \ArrC{\Pa}{\Pc}{30}{150} \CoilC{\Pa}{\Pc}{330}{210} }
	+ \tz{ \focus \PointC{3} \Arr{\Pa}{\Pc} } \bigg]
	\notag
	\\
	&-\rho^2(w_3-w_2) \bigg[ \tz{ \focus \PointB{3} \PointD{4} \Arr{\Pa}{\Pb} \Sol{\Pa}{\Pd} \Coil{\Pb}{\Pd} }
	+
	\tz{ \focus \PointB{3} \PointD{4} \Arr{\Pa}{\Pb} \SolC{\Pa}{\Pd}{0}{140} \CoilC{\Pa}{\Pd}{300}{180} }\bigg]
	\notag
	\\
	&- \rho^3(w_4-2w_3+w_2)
	\tz{\focus \PointB{3} \PointC{5} \PointD{4} \Arr{\Pa}{\Pb} \Sol{\Pa}{\Pc} \Coil{\Pb}{\Pc} \SolC{\Pa}{\Pd}{0}{140} \CoilC{\Pa}{\Pd}{300}{180} }
	+\frac{\sigma^2}{2} \partial_{\phi_1}^2 p(\phi_1)
	\label{eq:pnormalized}
\end{align}
Here, $w_k$ denotes the expectation value of the weight function for $k+l$ neighbors with respect to the number of neighbor distribution $q_n(l)$, that is
\begin{align}
	w_k := \langle w(k+l)\rangle = \sum_{l=0}^{\infty} w(k+l) q_n(l).
	\label{eq:neighbor_distr_expectation}
\end{align}
The number of neighbor distribution for correlated particles was recently derived in \cite{KSZI20}.
In case that only two-particle correlations are present this distribution depends on the two parameters
\begin{align}
	C_2:=& N^2 \int_{}^{} G_2(1,2) \theta_{1} \theta_2 \diff 1 \diff 2,
	\notag
	\\
	D_2:=& N \int_{}^{} G_2(1,2) \theta_{12} \diff 1 \diff 2,
	\label{eq:correlation_coefficients}
\end{align}
where
\begin{align}
	\theta_i:= \theta(R-\sqrt{x_i^2+y_i^2})
	\label{eq:thetai}
\end{align}
and $\theta_{ij}$ is given by Eq.~\eqref{eq:thetaij} with the Heaviside function $\theta$.
The number of neighbor distribution is given in \cite{KSZI20} as
\begin{align}
	q_n(l)=q(l)-[q(l-1)-q(l)](D_2-C1),
	\label{eq:poissonlike1}
\end{align}
where $q(l)$ is the following infinite series
\begin{align}
	q(l)=& 
	(C_1-C_2)^l \exp(C_2/2-C_1)
	\notag
	\\
	&\times \sum_{k=0}^{\infty} \bigg[ \frac{C_2}{2(C_1-C_2)^2}\bigg]^{k}
	 \frac{1}{k!(l-2k)!}.
	\label{eq:poissonlike2}
\end{align}
As before, $C_1=\frac{N}{L^2}\pi R^2$ is just the average number of neighbors.
Alternatively to the representation by an infinite sum \eqref{eq:poissonlike2} one can give a simple expression for the characteristic function of $q(l)$ \cite{KSZI20}
\begin{align}
    \chi(u)= \exp\bigg[ \sum_{l=1}^{2} \sum_{t=0}^l (-1)^{l+t} \frac{C_l}{l!} \binom{l}{t} \exp(itu) \bigg].
    \label{eq:characteristicfunctiongeneral}
\end{align}{}
Coming back to Eq.~\eqref{eq:pnormalized}, we give a guide to explain all interaction integrals.
Those are all possible diagrams where each particle is connected to $1$ via a solid line or an arrow (that means all particles are neighbors of particle $1$) such that every point is connected to $1$ also via a path consisting only of arrows and/or coils and there is exactly one arrow that starts from $1$. 
Particles that are not connected to $1$ via a path of arrows and/or coils do not directly take part in the interactions and thus do not appear in the diagrams.
Because there is only one arrow in the diagrams and each point can be connected to only one coil (that represents $g_2$) the diagrams can have no more than four points.
Each diagram has a prefactor of $-\rho^{k-1}$, where $k$ is the number of particles involved in the diagram. 
This prefactor is a combination of a combinatorial factor and $1/L^{2k}$ from Eqs.~\eqref{eq:homogeneousP} or \eqref{eq:reduced_gtwo}.
Furthermore, there is another prefactor of $w_{k}$ that takes into account the expectation value of the weight function when an average over all particles not involved in the diagram is performed. 
In principle, we would have additional diagrams where points not connected via an arrow are connected to $1$ via dashed lines, that means those particles are not a neighbor of particle $1$.
However, we can replace each dashed line by a solid line and multiply with $-1$ to compensate it due to property \eqref{eq:correlationintegralsubset}.
However, for these additional diagrams we have different prefactors of $w_{k-1}$ or $w_{k-2}$ if one or two dashed lines are involved, respectively, due to the different number of neighbors of particle $1$.
Considering all the aforementioned interaction terms, angular diffusion and streaming we arrive at Eq.~\eqref{eq:pnormalized}
Note, that for pair interactions, $w(n)=const.$, we have $w_2=w_3=w_4$ and only the first two interaction integrals remain.

Analogously to Eq.~\eqref{eq:pnormalized}, we find starting from Eq.~\eqref{eq:FPfull}, the time evolution equation for the two particle probability distribution
\begin{widetext}
\begin{align}
	&\partial_t p_2(\phi_1, \phi_2, \Delta)= -w_2 \bigg[\tz{\focus \PointC{2} \ArrC{\Pa}{\Pc}{30}{150} \CoilC{\Pa}{\Pc}{330}{210}} + \tz{\focus \PointC{2} \Arr{\Pa}{\Pc}}\bigg] 
	\notag
	\\
	&-\rho w_3 \bigg[
	\tz{\focus \PointB{2} \PointC{3} \Arr{\Pa}{\Pc} \Sol{\Pa}{\Pb}}
	+\tz{\focus \PointB{2} \PointC{3} \Arr{\Pa}{\Pc} \Coil{\Pb}{\Pc}\Sol{\Pa}{\Pb}}
	+\tz{\focus \PointB{2} \PointC{3} \ArrC{\Pa}{\Pc}{30}{150} \CoilC{\Pa}{\Pc}{330}{210}\Sol{\Pa}{\Pb}}
	+\tz{\focus \PointB{2} \PointC{3} \Arr{\Pa}{\Pc} \CoilC{\Pa}{\Pb}{60}{180} \SolC{\Pa}{\Pb}{0}{240}}
	\bigg]
	\notag
	\\
	&-\rho^2( w_4 -w_3) \bigg[
	\tz{\focus \PointB{2} \PointC{3} \PointD{4} \Arr{\Pa}{\Pc} \CoilC{\Pa}{\Pb}{60}{180} \Coil{\Pc}{\Pd} \SolC{\Pa}{\Pb}{0}{240} \Sol{\Pa}{\Pd}}
	+\tz{\focus \PointB{2} \PointC{3} \PointD{4} \Arr{\Pa}{\Pc}  \Coil{\Pc}{\Pd} \Sol{\Pa}{\Pb} \Sol{\Pa}{\Pd}}
	+\tz{\focus \PointB{2} \PointC{3} \PointD{4} \Arr{\Pa}{\Pc} \CoilC{\Pa}{\Pd}{350}{140} \Coil{\Pb}{\Pc} \Sol{\Pa}{\Pb} \SolC{\Pa}{\Pd}{300}{180}}
	+\tz{\focus \PointB{2} \PointC{3} \PointD{4} \Arr{\Pa}{\Pc} \CoilC{\Pa}{\Pd}{350}{140} \Sol{\Pa}{\Pb} \SolC{\Pa}{\Pd}{300}{180}}
	+\tz{\focus \PointD{2} \PointC{3} \PointE{4} \ArrC{\Pa}{\Pc}{30}{150} \CoilC{\Pa}{\Pc}{330}{210} \Coil{\Pd}{\Pe} \Sol{\Pa}{\Pd} \Sol{\Pa}{\Pe}}
	+\tz{\focus \PointD{2} \PointC{3} \PointE{4} \Arr{\Pa}{\Pc} \Coil{\Pd}{\Pe} \Sol{\Pa}{\Pd} \Sol{\Pa}{\Pe}}
	\bigg]
	\notag
	\\
	&-\rho^3 ( w_5 -2 w_4 +w_3) \bigg[
	\tz{\focus \PointB{3} \PointC{2} \PointD{5} \PointE{4} \Arr{\Pa}{\Pb} \CoilC{\Pa}{\Pe}{300}{170} \Coil{\Pc}{\Pd} \Sol{\Pa}{\Pc} \Sol{\Pa}{\Pd} \SolC{\Pa}{\Pe}{330}{100}}
	+\tz{\focus \PointB{3} \PointC{5} \PointD{2} \PointE{4} \Arr{\Pa}{\Pb} \CoilC{\Pa}{\Pe}{300}{170} \Coil{\Pb}{\Pc} \Sol{\Pa}{\Pd} \SolC{\Pa}{\Pe}{330}{100} \Sol{\Pa}{\Pc}}
	+\tz{\focus \PointB{3} \PointC{5} \PointD{2} \PointE{4} \Arr{\Pa}{\Pb} \Coil{\Pd}{\Pe} \Coil{\Pb}{\Pc} \Sol{\Pa}{\Pd} \Sol{\Pa}{\Pc} \Sol{\Pa}{\Pe}}
	\bigg]
	-\rho^4 ( w_6 -3 w_5 + 3 w_4 -w_3) \bigg[
	\tz{\focus \PointB{3} \PointC{6} \PointD{4} \PointE{2} \PointF{5} \Arr{\Pa}{\Pb} \CoilC{\Pa}{\Pd}{350}{140} \Coil{\Pe}{\Pf} \Coil{\Pb}{\Pc} \Sol{\Pa}{\Pf} \Sol{\Pa}{\Pc} \Sol{\Pa}{\Pe} \SolC{\Pa}{\Pd}{310}{180} }
	\bigg]
	\notag
	\\
	&-\rho w_2 \bigg[
	\tz{\focus \PointB{2} \PointC{3} \Arr{\Pa}{\Pc} \Dash{\Pa}{\Pb}}
	+\tz{\focus \PointB{2} \PointC{3} \Arr{\Pa}{\Pc} \Coil{\Pb}{\Pc}\Dash{\Pa}{\Pb}}
	+\tz{\focus \PointB{2} \PointC{3} \ArrC{\Pa}{\Pc}{30}{150} \CoilC{\Pa}{\Pc}{330}{210}\Dash{\Pa}{\Pb}}
	+\tz{\focus \PointB{2} \PointC{3} \Arr{\Pa}{\Pc} \CoilC{\Pa}{\Pb}{60}{180} \DashC{\Pa}{\Pb}{0}{240}}
	\bigg]
	\notag
	\\
	&-\rho^2( w_3 -w_2) \bigg[
	\tz{\focus \PointB{2} \PointC{3} \PointD{4} \Arr{\Pa}{\Pc} \CoilC{\Pa}{\Pb}{60}{180} \Coil{\Pc}{\Pd} \DashC{\Pa}{\Pb}{0}{240} \Sol{\Pa}{\Pd}}
	+\tz{\focus \PointB{2} \PointC{3} \PointD{4} \Arr{\Pa}{\Pc}  \Coil{\Pc}{\Pd} \Dash{\Pa}{\Pb} \Sol{\Pa}{\Pd}}
	+\tz{\focus \PointB{2} \PointC{3} \PointD{4} \Arr{\Pa}{\Pc} \CoilC{\Pa}{\Pd}{350}{140} \Coil{\Pb}{\Pc} \Dash{\Pa}{\Pb} \SolC{\Pa}{\Pd}{300}{180}}
	+\tz{\focus \PointB{2} \PointC{3} \PointD{4} \Arr{\Pa}{\Pc} \CoilC{\Pa}{\Pd}{350}{140} \Dash{\Pa}{\Pb} \SolC{\Pa}{\Pd}{300}{180}}
	+\tz{\focus \PointD{2} \PointC{3} \PointE{4} \ArrC{\Pa}{\Pc}{30}{150} \CoilC{\Pa}{\Pc}{330}{210} \Coil{\Pd}{\Pe} \Dash{\Pa}{\Pd} \Sol{\Pa}{\Pe}}
	+\tz{\focus \PointD{2} \PointC{3} \PointE{4} \Arr{\Pa}{\Pc} \Coil{\Pd}{\Pe} \Dash{\Pa}{\Pd} \Sol{\Pa}{\Pe}}
	\bigg]
	\notag
	\\
	&-\rho^3 ( w_4 -2 w_3 +w_2) \bigg[
	\tz{\focus \PointB{3} \PointC{2} \PointD{5} \PointE{4} \Arr{\Pa}{\Pb} \CoilC{\Pa}{\Pe}{300}{170} \Coil{\Pc}{\Pd} \Dash{\Pa}{\Pc} \Sol{\Pa}{\Pd} \SolC{\Pa}{\Pe}{330}{100}}
	+\tz{\focus \PointB{3} \PointC{5} \PointD{2} \PointE{4} \Arr{\Pa}{\Pb} \CoilC{\Pa}{\Pe}{300}{170} \Coil{\Pb}{\Pc} \Dash{\Pa}{\Pd} \SolC{\Pa}{\Pe}{330}{100} \Sol{\Pa}{\Pc}}
	+\tz{\focus \PointB{3} \PointC{5} \PointD{2} \PointE{4} \Arr{\Pa}{\Pb} \Coil{\Pd}{\Pe} \Coil{\Pb}{\Pc} \Dash{\Pa}{\Pd} \Sol{\Pa}{\Pc} \Sol{\Pa}{\Pe}}
	\bigg]
	-\rho^4 ( w_5 -3 w_4 + 3 w_3 -w_2) \bigg[
	\tz{\focus \PointB{3} \PointC{6} \PointD{4} \PointE{2} \PointF{5} \Arr{\Pa}{\Pb} \CoilC{\Pa}{\Pd}{350}{140} \Coil{\Pe}{\Pf} \Coil{\Pb}{\Pc} \Sol{\Pa}{\Pf} \Sol{\Pa}{\Pc} \Dash{\Pa}{\Pe} \SolC{\Pa}{\Pd}{310}{180} }
	\bigg]
	\notag
	\\
	&+\frac{\sigma^2}{2}\partial_{\phi_1}^2 [p(\phi_1) p(\phi_2)+ g_2(\phi_1, \phi_2, \Delta)] + 1\leftrightarrow 2 
	\notag
	\\
	&+ v(\cos \phi_1 - \cos \phi_2) \partial_{\Delta_x}g_2(\phi_1, \phi_2, \Delta) + v(\sin \phi_1 - \sin \phi_2) \partial_{\Delta_y}g_2(\phi_1, \phi_2, \Delta),
	\label{eq:p2}
\end{align}
\end{widetext}
where $1\leftrightarrow 2$ is an abbreviation for all previous terms with particles one and two interchanged.

Employing Eq.~\eqref{eq:g2}, in our notation \eqref{eq:homogeneousP}, \eqref{eq:reduced_ptwo} and \eqref{eq:reduced_gtwo},
\begin{align}
	g_2(\phi_1,\phi_2, \Delta)= p_2(\phi_1, \phi_2, \Delta) - L^2 p_{\phi}(\phi_1)p_{\phi}(\phi_2),
	\label{eq:reduced_ptwo2}
\end{align}
we obtain the time evolution equation of the pair correlation function
\begin{align}
	\partial_t g_2(\phi_1,\phi_2, \Delta)=& \partial_t p_2(\phi_1, \phi_2, \Delta) - L^2 p_{\phi}(\phi_1) \partial_t p_{\phi}(\phi_2) 
\notag
\\
&- L^2 p_{\phi}(\phi_2) \partial_t p_{\phi}(\phi_1).
\label{eq:time_evolution_g2_definition}
\end{align}
Inserting Eqs.~\eqref{eq:pnormalized} and \eqref{eq:p2} yields
\begin{widetext}
\begin{align}
	&\partial_t g_2(\phi_1, \phi_2, \Delta) =-w_2 \bigg[\tz{\focus \PointC{2} \ArrC{\Pa}{\Pc}{30}{150} \CoilC{\Pa}{\Pc}{330}{210}} + \tz{\focus \PointC{2} \Arr{\Pa}{\Pc}}\bigg] 
	\notag
	-\rho w_3 \bigg[
	\tz{\focus \PointB{2} \PointC{3} \Arr{\Pa}{\Pc} \Coil{\Pb}{\Pc}\Sol{\Pa}{\Pb}}
	+\tz{\focus \PointB{2} \PointC{3} \Arr{\Pa}{\Pc} \CoilC{\Pa}{\Pb}{60}{180} \SolC{\Pa}{\Pb}{0}{240}}
	\bigg]
	\notag \\
	&-\rho (w_3-w_2) \bigg[
	\tz{\focus \PointB{2} \PointC{3} \Arr{\Pa}{\Pc} \Sol{\Pa}{\Pb}}
	+\tz{\focus \PointB{2} \PointC{3} \ArrC{\Pa}{\Pc}{30}{150} \CoilC{\Pa}{\Pc}{330}{210}\Sol{\Pa}{\Pb}}
	\bigg]
	-\rho^2( w_4 -w_3) \bigg[
	\tz{\focus \PointB{2} \PointC{3} \PointD{4} \Arr{\Pa}{\Pc} \CoilC{\Pa}{\Pb}{60}{180} \Coil{\Pc}{\Pd} \SolC{\Pa}{\Pb}{0}{240} \Sol{\Pa}{\Pd}}
	+\tz{\focus \PointB{2} \PointC{3} \PointD{4} \Arr{\Pa}{\Pc} \CoilC{\Pa}{\Pd}{350}{140} \Coil{\Pb}{\Pc} \Sol{\Pa}{\Pb} \SolC{\Pa}{\Pd}{300}{180}}
	+\tz{\focus \PointD{2} \PointC{3} \PointE{4} \ArrC{\Pa}{\Pc}{30}{150} \CoilC{\Pa}{\Pc}{330}{210} \Coil{\Pd}{\Pe} \Sol{\Pa}{\Pd} \Sol{\Pa}{\Pe}}
	+\tz{\focus \PointD{2} \PointC{3} \PointE{4} \Arr{\Pa}{\Pc} \Coil{\Pd}{\Pe} \Sol{\Pa}{\Pd} \Sol{\Pa}{\Pe}}
	\bigg]
	\notag
	\\
	&-\rho^2( w_4 -2w_3+w_2) \bigg[
	\tz{\focus \PointB{2} \PointC{3} \PointD{4} \Arr{\Pa}{\Pc}  \Coil{\Pc}{\Pd} \Sol{\Pa}{\Pb} \Sol{\Pa}{\Pd}}
	+\tz{\focus \PointB{2} \PointC{3} \PointD{4} \Arr{\Pa}{\Pc} \CoilC{\Pa}{\Pd}{350}{140} \Sol{\Pa}{\Pb} \SolC{\Pa}{\Pd}{300}{180}}
	\bigg]
	-\rho^3 ( w_5 -2 w_4 +w_3) \bigg[
	\tz{\focus \PointB{3} \PointC{2} \PointD{5} \PointE{4} \Arr{\Pa}{\Pb} \CoilC{\Pa}{\Pe}{300}{170} \Coil{\Pc}{\Pd} \Sol{\Pa}{\Pc} \Sol{\Pa}{\Pd} \SolC{\Pa}{\Pe}{330}{100}}
	+\tz{\focus \PointB{3} \PointC{5} \PointD{2} \PointE{4} \Arr{\Pa}{\Pb} \Coil{\Pd}{\Pe} \Coil{\Pb}{\Pc} \Sol{\Pa}{\Pd} \Sol{\Pa}{\Pc} \Sol{\Pa}{\Pe}}
	\bigg]
	\notag
	\\
	&-\rho^3 ( w_5 -3 w_4 +3 w_3 -w_2) \bigg[
	\tz{\focus \PointB{3} \PointC{5} \PointD{2} \PointE{4} \Arr{\Pa}{\Pb} \CoilC{\Pa}{\Pe}{300}{170} \Coil{\Pb}{\Pc} \Sol{\Pa}{\Pd} \SolC{\Pa}{\Pe}{330}{100} \Sol{\Pa}{\Pc}}
	\bigg]
	-\rho^4 ( w_6 -3 w_5 + 3 w_4 -w_3) \bigg[
	\tz{\focus \PointB{3} \PointC{6} \PointD{4} \PointE{2} \PointF{5} \Arr{\Pa}{\Pb} \CoilC{\Pa}{\Pd}{350}{140} \Coil{\Pe}{\Pf} \Coil{\Pb}{\Pc} \Sol{\Pa}{\Pf} \Sol{\Pa}{\Pc} \Sol{\Pa}{\Pe} \SolC{\Pa}{\Pd}{310}{180} }
	\bigg]
	\\
	\notag
	&-\rho w_2 \bigg[
	\tz{\focus \PointB{2} \PointC{3} \Arr{\Pa}{\Pc} \Coil{\Pb}{\Pc}\Dash{\Pa}{\Pb}}
	+\tz{\focus \PointB{2} \PointC{3} \Arr{\Pa}{\Pc} \CoilC{\Pa}{\Pb}{60}{180} \DashC{\Pa}{\Pb}{0}{240}}
	\bigg]
	-\rho^2( w_3 -w_2) \bigg[
	\tz{\focus \PointB{2} \PointC{3} \PointD{4} \Arr{\Pa}{\Pc} \CoilC{\Pa}{\Pb}{60}{180} \Coil{\Pc}{\Pd} \DashC{\Pa}{\Pb}{0}{240} \Sol{\Pa}{\Pd}}
	+\tz{\focus \PointB{2} \PointC{3} \PointD{4} \Arr{\Pa}{\Pc} \CoilC{\Pa}{\Pd}{350}{140} \Coil{\Pb}{\Pc} \Dash{\Pa}{\Pb} \SolC{\Pa}{\Pd}{300}{180}}
	+\tz{\focus \PointD{2} \PointC{3} \PointE{4} \ArrC{\Pa}{\Pc}{30}{150} \CoilC{\Pa}{\Pc}{330}{210} \Coil{\Pd}{\Pe} \Dash{\Pa}{\Pd} \Sol{\Pa}{\Pe}}
	+\tz{\focus \PointD{2} \PointC{3} \PointE{4} \Arr{\Pa}{\Pc} \Coil{\Pd}{\Pe} \Dash{\Pa}{\Pd} \Sol{\Pa}{\Pe}}
	\bigg]
	\\
	\notag
	&-\rho^3 ( w_4 -2 w_3 +w_2) \bigg[
	\tz{\focus \PointB{3} \PointC{2} \PointD{5} \PointE{4} \Arr{\Pa}{\Pb} \CoilC{\Pa}{\Pe}{300}{170} \Coil{\Pc}{\Pd} \Dash{\Pa}{\Pc} \Sol{\Pa}{\Pd} \SolC{\Pa}{\Pe}{330}{100}}
	+\tz{\focus \PointB{3} \PointC{5} \PointD{2} \PointE{4} \Arr{\Pa}{\Pb} \Coil{\Pd}{\Pe} \Coil{\Pb}{\Pc} \Dash{\Pa}{\Pd} \Sol{\Pa}{\Pc} \Sol{\Pa}{\Pe}}
	\bigg]
	-\rho^4 ( w_5 -3 w_4 + 3 w_3 -w_2) \bigg[
	\tz{\focus \PointB{3} \PointC{6} \PointD{4} \PointE{2} \PointF{5} \Arr{\Pa}{\Pb} \CoilC{\Pa}{\Pd}{350}{140} \Coil{\Pe}{\Pf} \Coil{\Pb}{\Pc} \Sol{\Pa}{\Pf} \Sol{\Pa}{\Pc} \Dash{\Pa}{\Pe} \SolC{\Pa}{\Pd}{310}{180} }
	\bigg]
	\notag
	\\
	&+\frac{\sigma^2}{2}\partial_{\phi_1}^2  g_2(\phi_1, \phi_2, \Delta) 
	+ 1\leftrightarrow 2 
	\notag
	\\
	&+ v(\cos \phi_1 - \cos \phi_2) \partial_{\Delta_x}g_2(\phi_1, \phi_2, \Delta) + v(\sin \phi_1 - \sin \phi_2) \partial_{\Delta_y}g_2(\phi_1, \phi_2, \Delta).
	\label{eq:g2_dynamics}
\end{align}
\end{widetext}
Note that, for example, the multiplication
\begin{align}
[\partial_t p_\phi(\phi_1)] \times p_\phi(\phi_2)= & [\partial_t p_\phi(\phi_1)] \tz{\PointC{2} \Sol{\Pa}{\Pc}} 
\notag 
\\
&+[\partial_t p_\phi(\phi_1)]\tz{\PointC{2} \Dash{\Pa}{\Pc}}
\label{eq:symbolic_multiplication}
\end{align}
can be done symbolically.
A number of collision integrals in $\partial_t p_2$ and $- L^2 p_{\phi}(\phi_1) \partial_t p_{\phi}(\phi_2)$ or $- L^2 p_{\phi}(\phi_2) \partial_t p_{\phi}(\phi_1)$ cancel such as for example $\tz{\focus \PointB{2} \PointC{3} \Arr{\Pa}{\Pc} \Dash{\Pa}{\Pb}}$. 
However, their counter parts with particles one and two being neighbors, such as for example $\tz{\focus \PointB{2} \PointC{3} \Arr{\Pa}{\Pc} \Sol{\Pa}{\Pb}}$ do not cancel.
This is an effect of the $N-$particle interactions. 
As a result these terms appear with different sign but also with different prefactors and thus they do not cancel.
This effect is responsible for a number of terms that are not present for pair interactions, $w(n)=const.$.

\section{Comparison of Ring-Kinetic Theory and Agent-Based Simulations\label{sec:numerics}}

\begin{figure}[h]
	\begin{center}
		\includegraphics{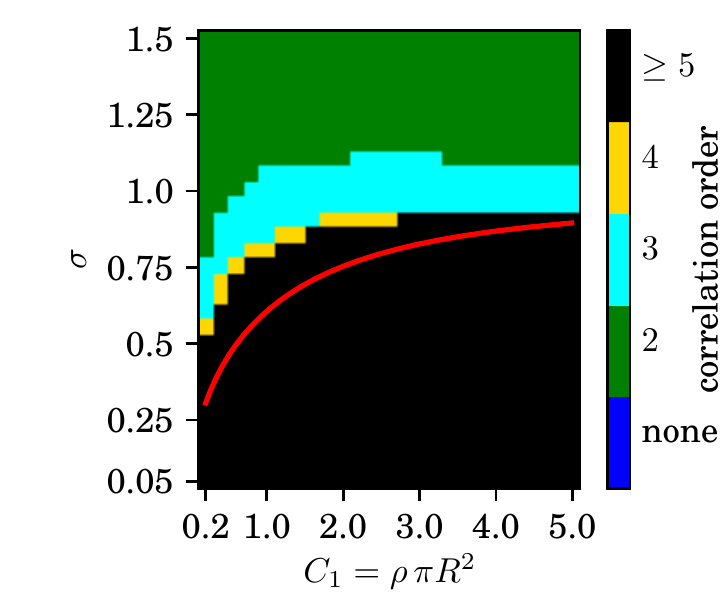}
	\end{center}
	\caption{Correlation map of the Vicsek-like model given by Eq.~\protect\eqref{eq:model} with $w(n)=1/n$.
	Colors encode the necessary order of correlations (e.g. cyan$\sim3\sim$three particle correlations, etc.) for a quantitative description of the system.
	The correlation map was obtained by a recently introduced method \protect\cite{KSZI20} comparing the measured number of neighbor distribution to a theoretically predicted one.
	One considers the distributions to be acceptably close if the corresponding Kullback-Leibler divergence is less than $10^{-3}$. 
	We simulated $24$ realizations for each noise strength $\sigma \in \{0.05, 0.1, 0.15, \dots, 1.5\}$ and each particle density $C_1=\rho\pi R^2\in\{0.2, 0.4, 0.6, \dots, 5.0 \}$ with parameters $v=R=1$, $N=10^4$. 
	The correlation parameters have been measured for a time interval of $10^4$ after a thermalization time of $10^3$.
	Eq.~\protect\eqref{eq:model} was integrated using an Euler-Maruyama scheme with time step $\Delta t = 10^{-2}$.
	The analysis has been done as in \protect\cite{KSZI20}.
	The red curve shows the mean field transition line given by Eq.~\protect\eqref{eq:sigmacmf2}.}
	\label{fig:correlation_map}
\end{figure}
Recently, a quantitative numerical method to predict the minimal required order of correlations has been introduced in \cite{KSZI20}.
It is based on the measurement of the number of neighbors distribution (neighbors are particles that are closer than $R$) of a randomly selected particle.
Besides this measurement, the number of neighbors distribution is also calculated under the assumption that only correlations up to a given order $l_{\text{max}}$ are present, that is $G_l\equiv 0$ for all $l>l_{\text{max}}$.
If both distributions agree reasonably well, the correlation order $l_{\text{max}}$ is considered to be sufficient, see \cite{KSZI20} for details.

In Fig.~\ref{fig:correlation_map} we show a correlation map that we obtained by this method for large systems of $N=10^4$ particles.
It shows, depending on parameters, which order of correlations is required.
We expect excellent quantitative predictions of the ring-kinetic theory in a parameter regime where pair correlations are sufficient.
Strictly speaking, this is the case in the disordered phase only.
However, we find that a reasonable ring-kinetic description of the system is still possible if the influence of higher order correlations is already measurable but still not dominant. 

We solve the time evolution equations for $p_{\phi}$ and $g_2$, Eqs.~\eqref{eq:pnormalized} and \eqref{eq:g2_dynamics}, numerically in Fourier space, see Appendix \ref{sec:fourier}. 
As a measurable quantity we consider the standard radial distribution function $g(r)$ that is defined by
\begin{align}
	g(r)= \frac{L^2}{N(N-1)}\sum_{i\neq j} \frac{1}{2\pi r} \langle \delta(r-|\mathbf{r}_i-\mathbf{r}_j|) \rangle.
	\label{eq:def_radial_distribution_function}
\end{align}
It is related to the correlation function $g_2$ given in Eq.~\eqref{eq:reduced_gtwo} by
\begin{align}
	g(r)= 1+ &\int_{0}^{2\pi}\diff \phi_1 \int_{0}^{2\pi} \diff \phi_2 \int_{0}^{L}\diff \Delta_x \int_{0}^{L} \diff \Delta_y
	\notag
	\\
	&\times g_2(\phi_1, \phi_2, \Delta)\frac{1}{2 \pi r} \delta(r-|\Delta|)
	\label{eq:relation_radial_distribution_function}
\end{align}
and can be calculated analytically from the Fourier representation, see Appendix \ref{sec:fourier}.

\begin{figure}[h]
	\begin{center}
		\includegraphics{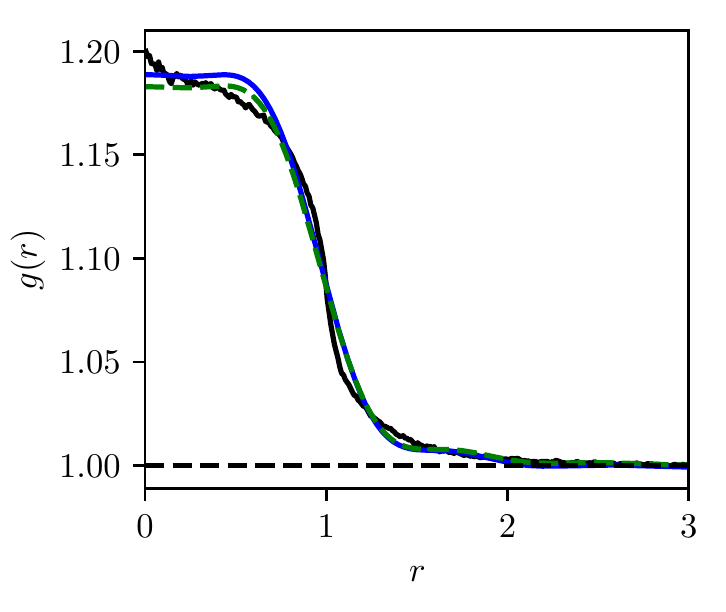}
	\end{center}
	\caption{Radial distribution function $g(r)$ obtained from ring-kinetic theory (blue solid line) and sampled directly from agent-based simulations according to Eq.~\protect\eqref{eq:def_radial_distribution_function} (black solid line).
	For the ring-kinetic theory we integrate the time evolution equations \protect\eqref{eq:pnormalized} and \protect\eqref{eq:g2_dynamics} in Fourier space (see Appendix \ref{sec:fourier}) with an Euler scheme with step size $\Delta t=10^{-2}$ for an absolute time of $100$, starting initially uncorrelated and disordered.
	For the agent based simulation we integrate Eq.~\protect\eqref{eq:model} with an Euler-Maruyama scheme with step size $\Delta t =10^{-2}$. 
	After a thermalization time of $10^3$ we average over a time of $10^4$ and over $24$ realizations.
	Parameters are $\sigma=1.5$, $v=R=1$, $N=509$, $C_1=\rho\pi R^2=1$, $L\approx 40$. 
	The ring-kinetic theory uses Fourier modes $F_{klmn}$ according to Eq.~\protect\eqref{eq:fourieransatzg1} with angular indexes $k,l \in \{-2, \dots,  2\}$ and spatial indexes $m,n \in \{-48, \dots, 48\}$.
	Performing a spatial Fourier transform of the measured curve (black line) with indexes from the same range $m,n\in \{-48, \dots, 48\}$ and transforming back into real space we obtain the dashed green line.
	The dashed solid line at $y=1$ serves as a guide to the eyes.}
	\label{fig:radial_distribution_function}
\end{figure}

In Fig.~\ref{fig:radial_distribution_function} we compare the radial distribution function measured in agent-based simulations with the results from ring-kinetic theory.
Overall they agree very well, however, there are minimal deviations.
Those deviations can be explained by the finite resolution used in the Fourier transform.
We evaluated the ring-kinetic equations in Fourier space using a minimal spatial wave length of $0.83$.
Since the radial distribution functions $g(r)$ drops rapidly at about $r\approx 1$ it is not perfectly resolved, causing small deviations.
To test this hypothesis, we Fourier transform the pair correlation function corresponding to the measured $g(r)$ with the same spatial resolution and Fourier transform it back into real space.
The resulting curve deviates by less than one percent from the ring-kinetic result, see Fig.~\ref{fig:radial_distribution_function}.

\begin{figure}[h]
	\begin{center}
		\includegraphics[width=0.95\columnwidth]{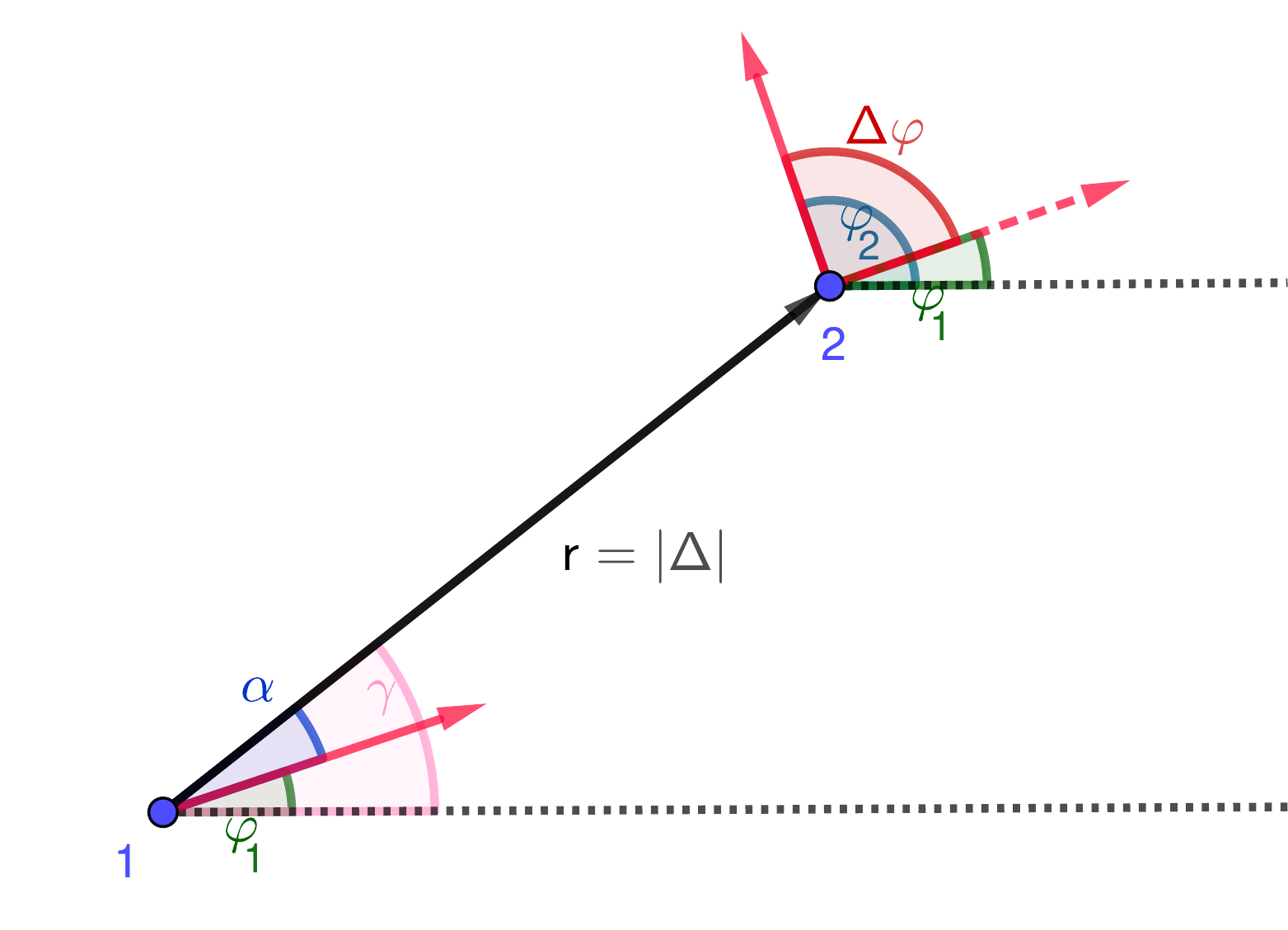}
	\end{center}
	\caption{Sketch of the three arguments $r$, $\alpha$ and $\Delta \phi$ of the pair correlation function $h$ for homogeneous and isotropic systems, see Eq.~\eqref{eq:definition_h_function} for the definition of $h(r, \alpha, \Delta \phi)$. 
	The blue points denote the positions of particles one and two. Their distance is denoted by $r$ and their direction of motion (red arrows) is given by the angles $\phi_1$ and $\phi_2$.
	The difference between those two angles is denoted by $\Delta \phi$.
	The difference between the polar angle $\gamma$ of the vector pointing from particle one to particle two and $\phi_1$ is denoted by $\alpha$.}
	\label{fig:sketch_parameters}
\end{figure}
The ring-kinetic theory predicts the full two particle correlation function $g_2(\phi_1, \phi_2, \Delta)$ and not only the radial distribution function $g(r)$ that is an integral of it, see Eq.~\eqref{eq:relation_radial_distribution_function}.
Without spontaneous symmetry breaking, that is in the disordered phase, the system is isotropic.
In that case, the pair correlation function depends only on three independent arguments and not on four, such as $\phi_1$, $\phi_2$, $\Delta_x$ and $\Delta_y$.
We choose those three degrees of freedom as the length of the vector $\Delta$, $r:=|\Delta|$, the difference between the orientations of the two particles, $\Delta\phi:=\phi_2-\phi_1$, and the difference of the polar angle of the vector $\Delta$ and the orientation of the first particle $\alpha:= \gamma-\phi_1$, where $\gamma$ is the polar angle of the vector $\Delta$, see Fig.~\ref{fig:sketch_parameters}.
Depending one those arguments we define the function
\begin{align}
	h(r, \alpha, \Delta \phi):=& 2\pi \int_{0}^{2\pi} \diff \phi_1 \int_{0}^{2\pi} \diff \phi_2 \int_{0}^{2\pi} \diff \gamma 
	\notag
	\\
	& \times g_2(\phi_1, \phi_2, \Delta_x=r\cos \gamma, \Delta_y=r\sin \gamma)
	\notag
	\\
	&\times \delta(\alpha-\gamma+\phi_1) \delta(\Delta \phi-\phi_2+\phi_1).
	\label{eq:definition_h_function}
\end{align}
For translational and rotational invariant systems it contains all information about two particle correlations.
It can be directly sampled from numerical data according to
\begin{align}
	&h(r, \alpha, \Delta \phi)= \frac{L^2 (2\pi)^2}{N(N-1)} \frac{1}{2\pi r}\sum_{i\neq j} \langle \delta(r-|\mathbf{r}_i-\mathbf{r}_j|) 
	\notag
	\\
	&\times \delta(\alpha-\gamma_{ij} + \phi_i) \delta(\Delta \phi -\phi_j+\phi_i) \rangle -1.
	\label{eq:sampling_h_function}
\end{align}

\subsection{Dependence of $h(r, \alpha, \Delta \phi)$ on $\Delta \phi$}

\begin{figure}[H]
	\begin{center}
		\includegraphics{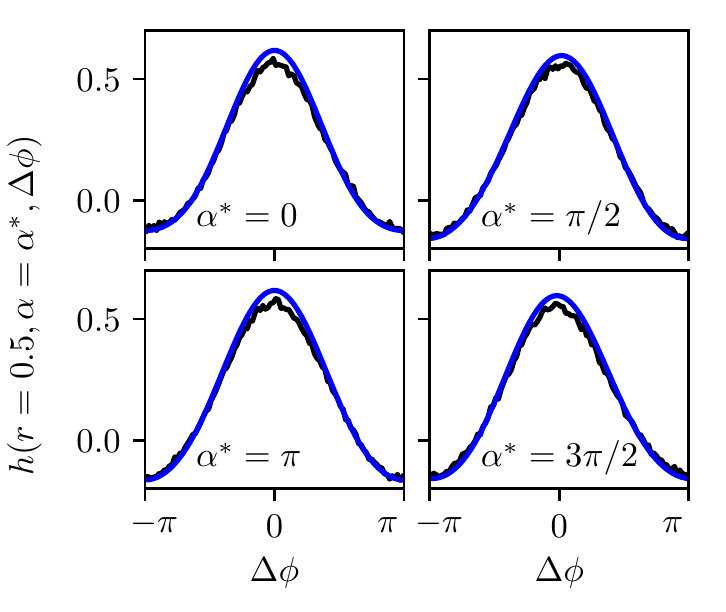}
	\end{center}
	\caption{Dependence of the correlation function $h(r, \alpha, \Delta \phi)$ on difference of velocity directions $\Delta \phi$ for fixed values of $r=0.5$ and $\alpha$ (value given in the plot).
	Results of the ring-kinetic theory (blue line) are compared to direct measurements of agent-based simulations (black line).
	Parameters as in Fig.~\protect\ref{fig:radial_distribution_function}.}
	\label{fig:h_of_deltaphi1}
\end{figure}

\begin{figure}[h]
	\begin{center}
		\includegraphics{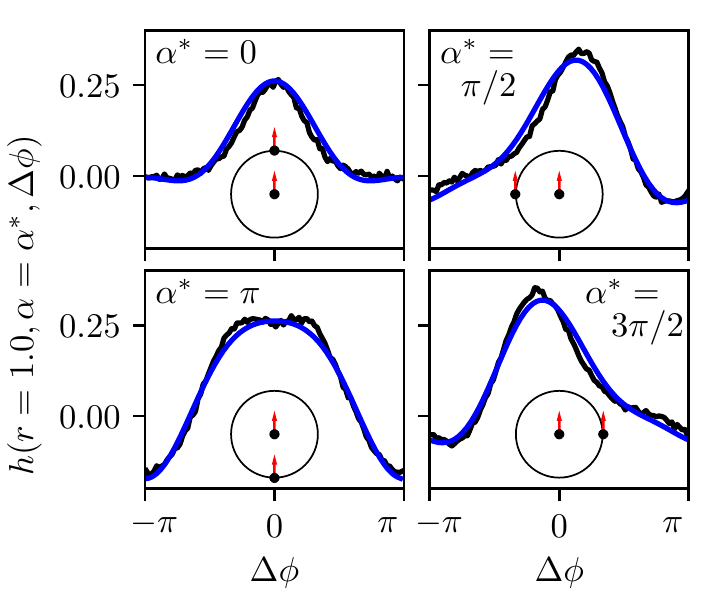}
	\end{center}
	\caption{Two particle correlation function $h$ as a function of the difference of the velocity directions of both particles as in Fig.~\protect\ref{fig:h_of_deltaphi1}, but here for $r=1$. The black bullets sketch the relative positions of particle one (in the center of the circle) and particle two (on the circumference of the circle). The red arrows indicate the direction of the particle velocities at $\Delta \phi=0$.}
	\label{fig:h_of_deltaphi2}
\end{figure}

We consider the dependence of the correlation function $h(r, \alpha, \Delta \phi)$ on the difference in orientation of the two particles $\Delta \phi$ for fixed $r$ and $\alpha$.
In Fig.~\ref{fig:h_of_deltaphi1} we fix $r=0.5$ and $\alpha = 0, \pi/2, \pi, 3\pi/2$.
That means that the two particles are in the interior of  each others interaction circle.
We see that the correlations are highest for $\Delta \phi=0$ showing that nearby particles align.
For $\Delta \phi=\pi$ the correlations are even slightly negative.
Rotations of one particle around the other seem to be of no particular importance at the considered distance as we see almost the same picture for different values of $\alpha$.
In Fig.~\ref{fig:h_of_deltaphi2} we fix $r=1.0$.
That means both particles are exactly at the boundary of each others interaction region.
For $\alpha=0$ and $\alpha=\pi$, that means if particle two is in the front or in the back of particle one (looking in the direction of motion of particle one), the distribution of $h(\Delta \phi)$ is still symmetric with a maximum at $\Delta \phi =0$.
If particle two is on the left or on the right of particle one it is slightly more likely that particle two points a bit outward the interaction region than to be perfectly aligned with particle one.
This is reasonable due to the following argument.
When particle two is placed at distance $r=1$ at a given time, it is more likely that it was at $r<1$ than that it was at $r>1$ shortly before, because the pair correlations are much higher for $r<1$.

\begin{figure}[h]
	\begin{center}
		\includegraphics{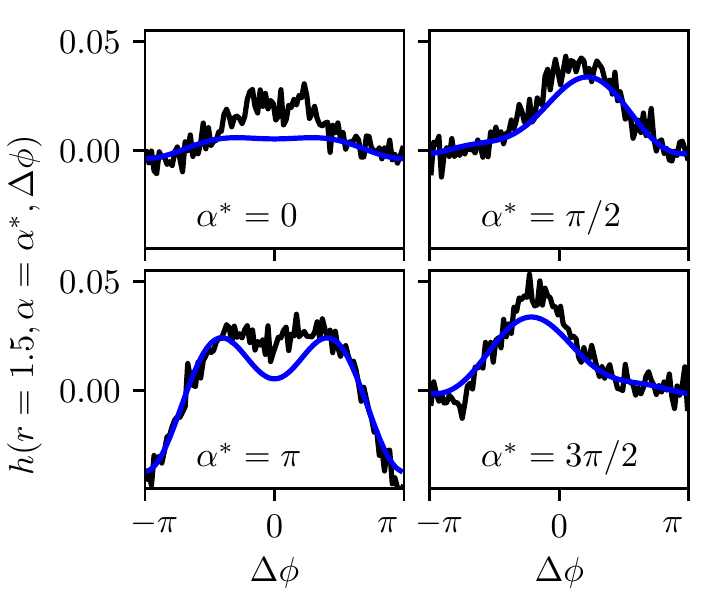}
	\end{center}
	\caption{Two particle correlation function $h$ as a function of the difference of the velocity directions of both particles as in Fig.~\protect\ref{fig:h_of_deltaphi1}, but here for $r=1.5$.}
	\label{fig:h_of_deltaphi3}
\end{figure}

In Fig.~\ref{fig:h_of_deltaphi3} we fix $r=1.5$.
Thus, the two particles are not directly interacting with each other.
We see that the strength of the correlations is much smaller in this case.
Also the fluctuations in the measurements of agent-based simulations are larger, because there are less events sampled.
Quantitatively, the correlations are similar to the case of $r=1.0$.
\subsection{Dependence of $h(r, \alpha, \Delta \phi)$ on $\alpha$}
\begin{figure}[H]
	\begin{center}
		\includegraphics{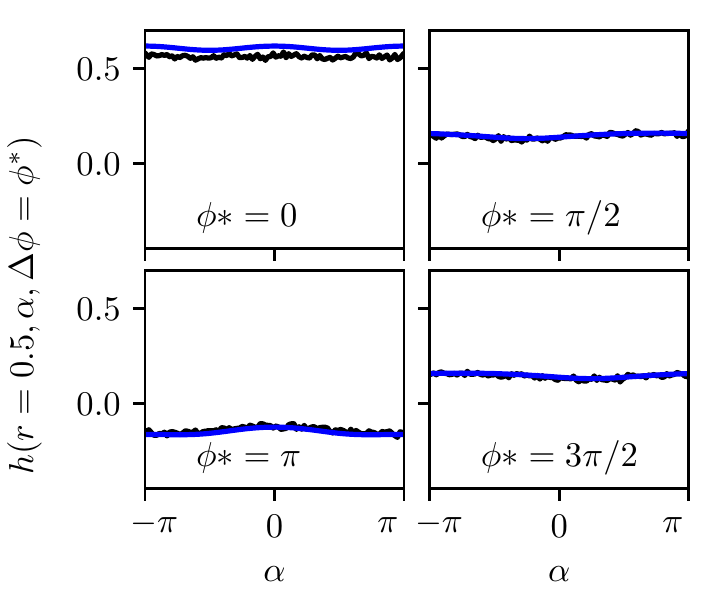}
	\end{center}
	\caption{Dependence of the correlation function $h(r, \alpha, \Delta \phi)$ on the angle $\alpha$ giving the rotation of particle two around particle with respect to the direction of motion of particle one. 
	Fixed values of $r=0.5$ and $\Delta \phi$ (value given in the plot) are considered.
	Results of the ring-kinetic theory (blue line) are compared to direct measurements of agent-based simulations (black line).
	Parameters as in Fig.~\protect\ref{fig:radial_distribution_function}.}
	\label{fig:h_of_alpha1}
\end{figure}

In Fig.~\ref{fig:h_of_alpha1} we consider the $\alpha$-dependence of the correlation function $h$ for $r=0.5$ and different values of $\Delta \phi$.
As discussed in the previous subsection, there is almost no $\alpha$-dependence.
The value of $h$ is largest for $\Delta \phi =0$ and smallest and even negative for $\Delta \phi = \pi$ which is in agreement with Fig.~\ref{fig:h_of_deltaphi1}.

\begin{figure}[h]
	\begin{center}
		\includegraphics{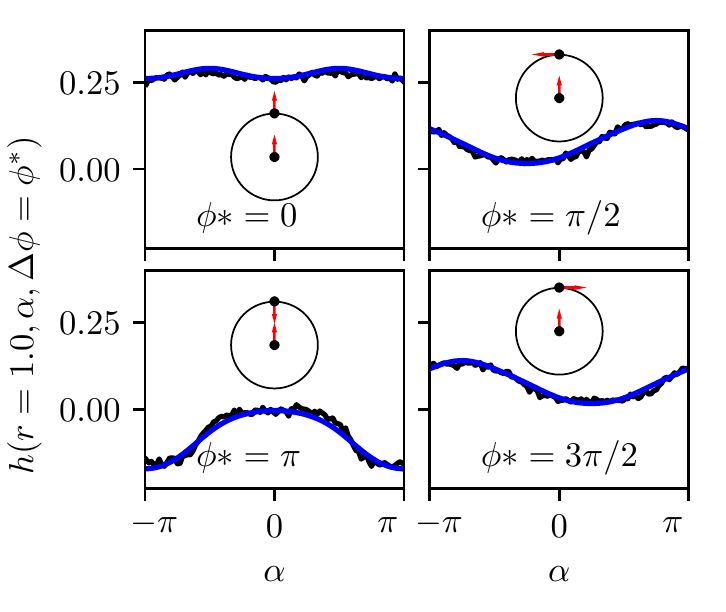}
	\end{center}
	\caption{Two particle correlation function $h$ as a function of the relative orientation $\alpha$ as in Fig.~\protect\ref{fig:h_of_alpha1}, but here for $r=1$. The black bullets sketch the relative positions of particle one (in the center of the circle) and particle two (on the circumference of the circle) at $\alpha=0$. The red arrows indicate the direction of the particle velocities.}
	\label{fig:h_of_alpha2}
\end{figure}

In Fig.~\ref{fig:h_of_alpha2} we fix $r=1$.
For aligned particles, $\Delta \phi =0$, the $\alpha$-dependence of the correlation function is still very small with a slight preference of particle two being right or left of particle one compared to a placement in the front or back of particle one.
For anti-aligned particles, $\Delta \phi= \pi$, it is unlikely that particle two is in the back of particle one.
In that case, the particles would move away from each other. 
That means that they have been closer to each other in the past which makes it unlikely that they are anti-aligned.
For $\Delta=\pi/2$ and $\Delta \phi =3\pi/2$, there is a complex $\alpha$-dependence with maximal correlations at $\alpha\approx 0.75 \pi$ and $\alpha\approx -0.75\pi$, respectively.
The value of $\alpha$ with minimal correlations is shifted by $\pi$ with respect to the values of maximal correlations.

\begin{figure}[h]
	\begin{center}
		\includegraphics{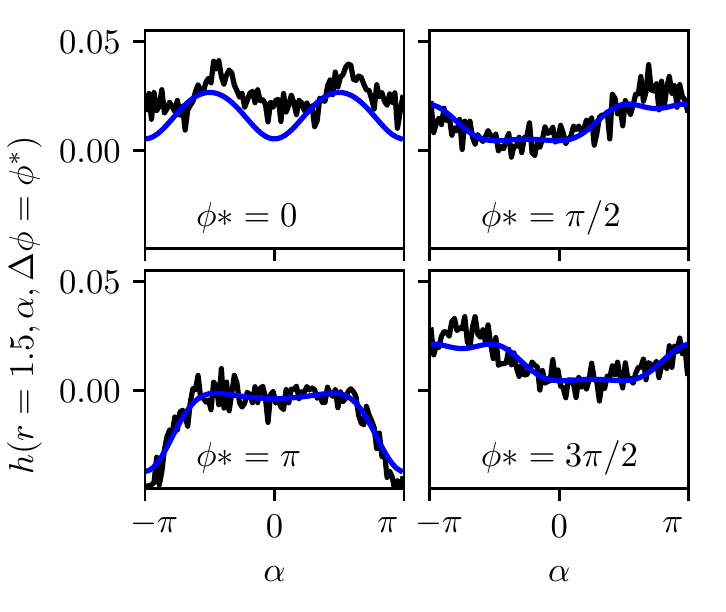}
	\end{center}
	\caption{Two particle correlation function $h$ as a function of the relative orientation $\alpha$ as in Fig.~\protect\ref{fig:h_of_alpha1}, but here for $r=1.5$.}
	\label{fig:h_of_alpha3}
\end{figure}

In Fig.~\ref{fig:h_of_alpha3} we fix $r=1.5$.
The correlation function behaves qualitatively similar to the case of $r=1$, however here, the correlations are much smaller.
\subsection{Dependence of $h(r, \alpha, \Delta \phi)$ on $r$}
\begin{figure}[H]
	\begin{center}
		\includegraphics{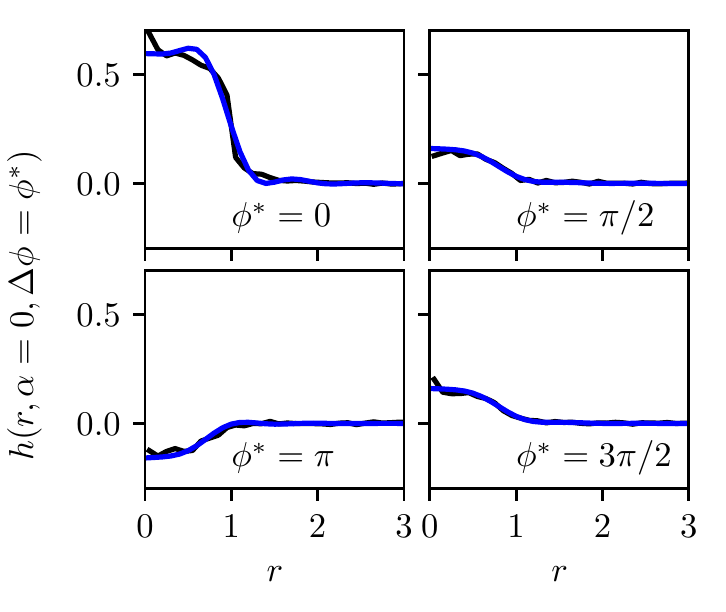}
	\end{center}
	\caption{Dependence of the correlation function $h(r, \alpha, \Delta \phi)$ on the distance $r$ for fixed values of $\alpha=0$ and $\Delta \phi$ (value given in the plot).
	Results of the ring-kinetic theory (blue line) are compared to direct measurements of agent-based simulations (black line).
	Parameters as in Fig.~\protect\ref{fig:radial_distribution_function}.}
	\label{fig:h_of_r1}
\end{figure}

\begin{figure}[h]
	\begin{center}
		\includegraphics{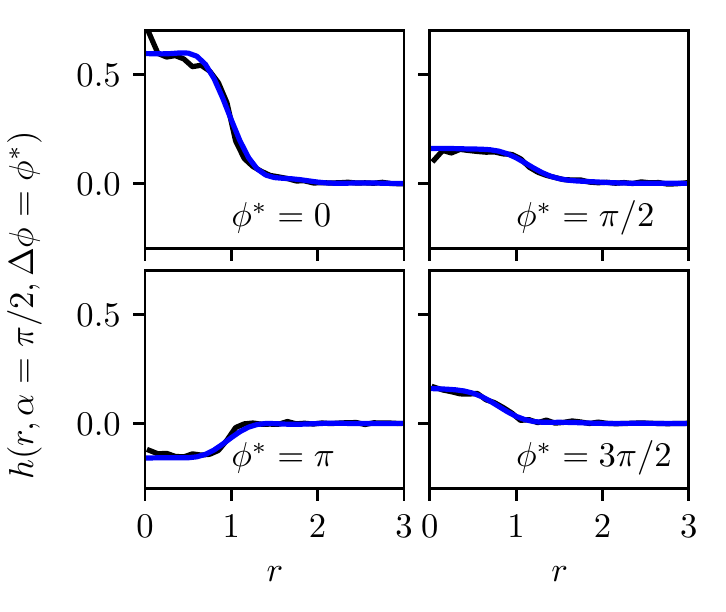}
	\end{center}
	\caption{Two particle correlation function $h$ as a function of distance $r$ as in Fig.~\protect\ref{fig:h_of_r1}, but here for $\alpha=\pi/2$.}
	\label{fig:h_of_r2}
\end{figure}

\begin{figure}[h]
	\begin{center}
		\includegraphics{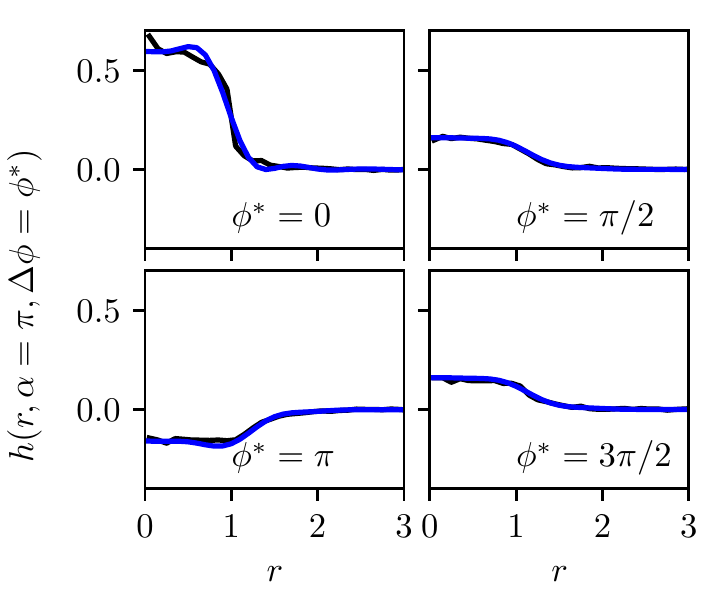}
	\end{center}
	\caption{Two particle correlation function $h$ as a function of distance $r$ as in Fig.~\protect\ref{fig:h_of_r1}, but here for $\alpha=\pi$.}
	\label{fig:h_of_r3}
\end{figure}

\begin{figure}[H]
	\begin{center}
		\includegraphics{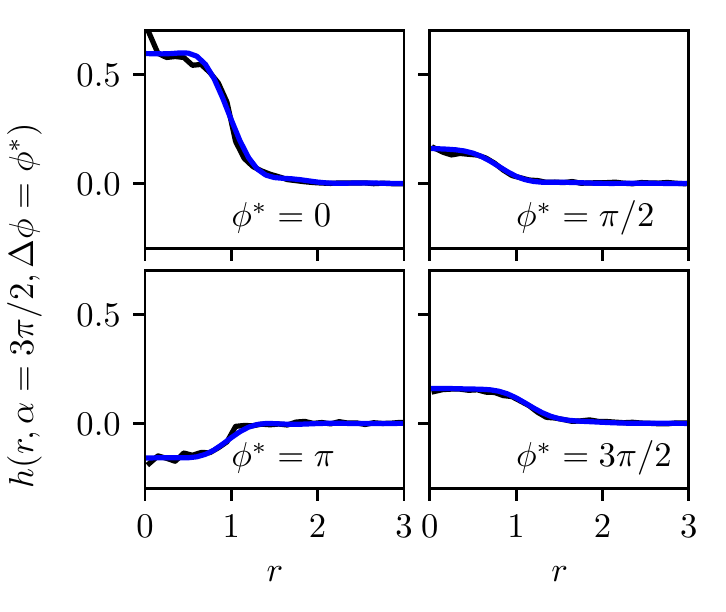}
	\end{center}
	\caption{Two particle correlation function $h$ as a function of distance $r$ as in Fig.~\protect\ref{fig:h_of_r1}, but here for $\alpha=3\pi/2$.}
	\label{fig:h_of_r4}
\end{figure}

In this subsection we consider the $r$-dependence of the correlation function $h$ for fixed values of $\alpha$ and $\Delta \phi$.
We find the same qualitative behavior for all considered values of $\alpha$, see Figs.~\ref{fig:h_of_r1}-\ref{fig:h_of_r4}.
For $\Delta \phi=0, \pi/2$ and $3\pi/2$, for small $r$ the function $h$ shows a plateau of high correlations and decreases relatively fast at about $r=1$ towards zero.
For $\Delta \phi = \pi$, the plateau for small $r$ is negative and the decay towards zero at about $r=1$ is still present.
This shows that nearby anti-aligned particles are unlikely.

\subsection{Applicability of the Ring-Kinetic Theory}

\begin{figure}[h]
	\begin{center}
		\includegraphics{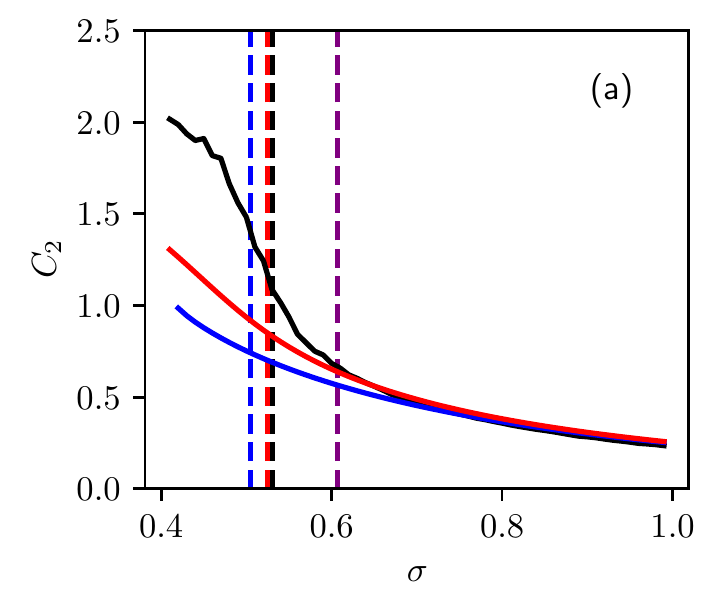}\\
		\includegraphics{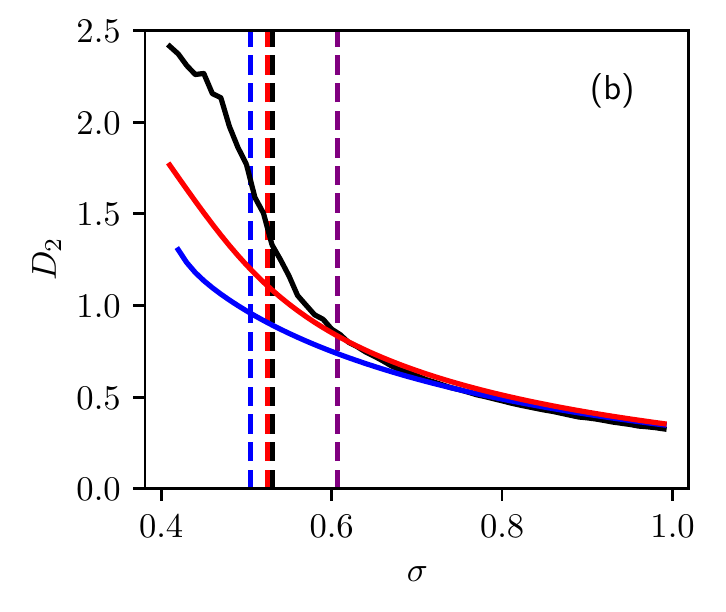}
	\end{center}
	\caption{Local spatial pair correlation parameters $C_2$ (a) and $D_2$ (b) as defined in Eq.~\protect\eqref{eq:correlation_coefficients} compared for direct simulations (black line), ring-kinetic theory (blue line), and a kinetic theory including a $g_3$-closure (red line).
	The dashed vertical lines shows the onset of flocking measured in agent-based simulations at $\sigma_{c, sim}=0.53$ (black), in the ring-kinetic theory at $\sigma_{c, rk}=0.505$ (blue), ring-kinetic theory with closure at $\sigma_{c, closure}=0.525$ (red) and in mean field theory at $\sigma_{c, mf}=0.607$ (purple).
	System parameters are: $C_1=\rho \pi R^2 =1$, $R=v=1$, $N=509$.}
	\label{fig:c2_d2}
\end{figure}

In the previous subsections we have seen that for the considered parameters, the ring-kinetic theory agrees very well with direct agent-based simulations.
There are only very small deviations partially caused by the finite resolution in the Fourier transform in the numerical implementation of the ring-kinetic equations.
In this subsection we study the applicability of the ring-kinetic theory depending on parameters.
We have seen in the previous subsections that the spatial pair correlations decay rather rapidly for $r \lesssim R$.
Therefore, we compare the integrated spatial pair correlations $C_2$ and $D_2$ defined in Eq.~\eqref{eq:correlation_coefficients} between ring-kinetic theory and direct simulations in Fig.~\ref{fig:c2_d2}.
We see very good agreement between ring-kinetic theory and simulations for noise strengths $\sigma \gtrsim 0.68$.
For smaller noise strengths we find serious deviations.

\begin{figure}[h]
	\begin{center}
		\includegraphics{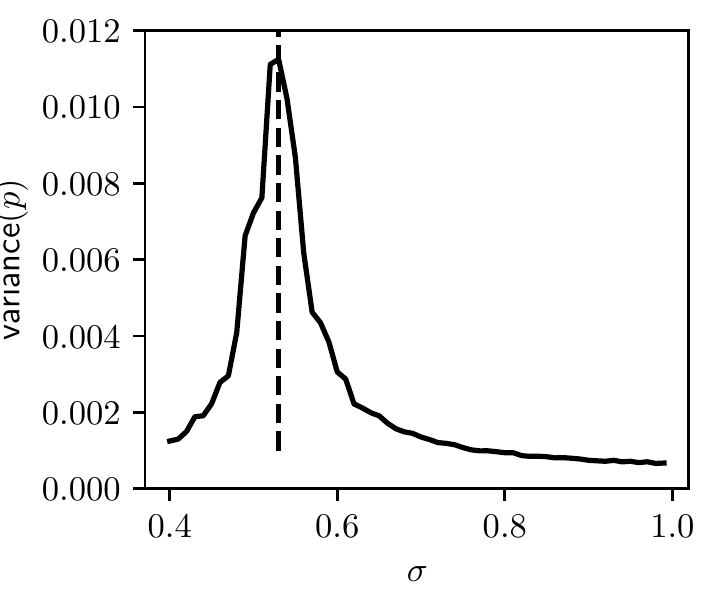}
	\end{center}
	\caption{Fluctuations of the polar order parameter $p$, measured in agent-based simulations. The maximum at $\sigma=0.53$ indicates the onset of collective motion.
	Parameters are as in Fig.~\protect\ref{fig:c2_d2}}
	\label{fig:fluctuations_polar_order}
\end{figure}

For the considered parameter set, homogeneous mean field theory, Eq.~\eqref{eq:sigmacmf2}, predicts the onset of collective motion at $\sigma_c\approx 0.607$.
In the direct simulations we determined the onset of flocking at $\sigma_{c, \text{sim}} = 0.53(1) $ from fluctuations of the polar order parameter $p=|\mathbf{p}|=\sqrt{p_x^2 + p_y^2}$, with $p_x= 1/N\sum_{i=1}^{N} \cos \phi_i$, $p_y= 1/N\sum_{i=1}^{N} \sin \phi_i$, see Fig.~\ref{fig:fluctuations_polar_order}.
\begin{figure}[h]
	\begin{center}
		\includegraphics{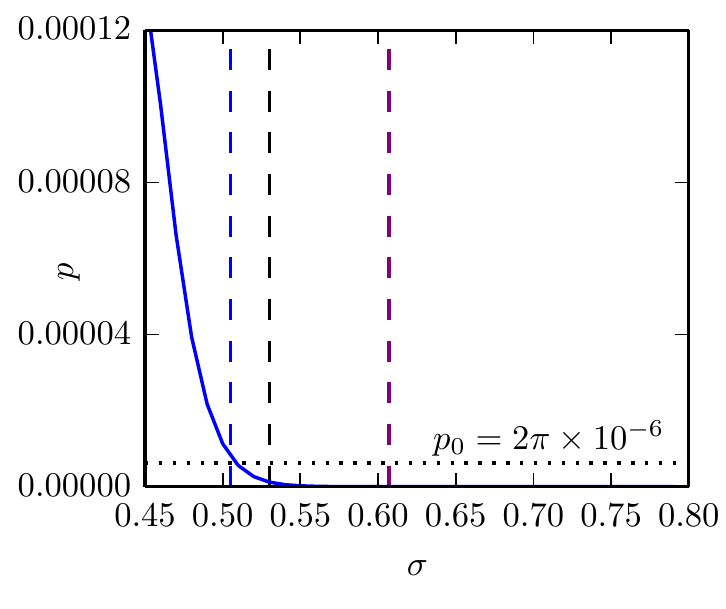}
	\end{center}
	\caption{Polar order parameter $p$ obtained with ring-kinetic theory.
The system was initiated without correlations and with a minimal polar order of $p_0=2\pi\times 10^{-6}\approx 6.3\times 10^{-6}$. The graph shows the polar order after a waiting time of $T=250$.
Far in the disordered regime the system is already in steady state after this time, that means the correlations are stationary. 
However, close to the transition and also in the polarly ordered regime the systems needs much longer to become stationary.
In particular at small noise, the polar order is still much below its steady state value.
Therefore, we estimate the transition by a numerical stability analysis of the disordered state: we consider the system as disordered if the polar order parameter after time $T$ is below its initial value $p_0$.
In that case we typically observe that the polar order parameter is still decreasing at the end of the observation time.
On the other hand we consider the system as ordered when the final value $p$ is larger than $p_0$.
In that case we also typically observe that $p$ is still increasing at the end of the observation time. 
The transition noise strength between polar order and disorder is displayed as the blue vertical dashed line at $\sigma=0.505$.
The black and purple vertical dashed lines show the transition noise strengths obtained in agent-based simulations and in mean field theory, respectively.
Parameters are as in Fig.~\protect\ref{fig:c2_d2}}
	\label{fig:polar_order_ringkinetic}
\end{figure}

Returning to the deviations of the spatial correlations predicted by ring-kinetic theory in Fig.~\ref{fig:c2_d2}, we find that they are significant already before the onset of flocking.
Nevertheless, the predicted correlations are not completely wrong, but have the correct order of magnitude.
Therefore, we can still obtain the onset of collective motion using the ring-kinetic theory.

In order to analyze fluctuations or susceptibilities we would require to analyze stationary states also within the ordered phase.
However, the solutions of the ring-kinetic equations require a very long time to become stationary in the ordered regime (when started with only a minimal polar order).
Therefore, we determine the onset of flocking within the ring-kinetic theory by a numerical stability analysis of the disordered state.
In the numerical solution of the ring-kinetic equations we start with some (very small) initial polar order because the disordered states are always (for small noise unstable) stationary solutions.
We started with an initial polar order of $p_0=2 \pi \times 10^{-6}$ and considered the state as polarly ordered if the final polar order (after a total time of $250$ or $1000$) is larger than $p_0$ and disordered if $p<p_0$ in the final state, see Fig.~\ref{fig:polar_order_ringkinetic}.
Typically we find also a decreasing trend of $p$ at the end of the observation time if $p<p_0$ and an increasing trend if $p>p_0$.
In that way, the ring-kinetic theory predicts the flocking transition at $\sigma_{c, \text{rk}}=0.505(5)$.
Surprisingly, that result is already pretty close to the value measured in direct simulations, even though the pair correlations are not quantitatively correct.
Apparently, considering correlations of the correct order of magnitude, improves mean field theory significantly.

Similar to phase transitions in equilibrium spin systems, correlations shift the flocking transition towards smaller noise, compared to mean field theory.
That means correlations favor disorder.
This can be understood qualitatively as follows.
If we consider a particle that moves not in the direction of the majority, without correlations it would be convinced to join the majority due to interactions with other particles (that on average behave as the majority).
If correlations are present, the interaction partners can have the same direction as the considered particle (different from the majority) due to correlations.
That means a particle that is oriented differently from the global average, is likely to be accompanied by other particles that differ from the global average if strong angular correlations are present.
This correlation effect weakens the alignment mechanism compared to mean field.

It should be mentioned that the considered system is relatively small ($N=509$).
The correlation map in Fig.~\ref{fig:correlation_map} is based on simulations of much larger systems ($N=10^4$).
In those larger systems, the flocking transition becomes discontinuous and higher order correlations or inhomogeneous solutions might become more important.
In fact, for larger systems, the transition noise strength is shifted towards larger noise strengths $\sigma_{c, \text{sim, large}}=0.58$ \cite{KSZI20}.
This effect can be understood by density fluctuations. 
Because the transition noise strength of the homogeneous solutions is highly density dependent, polar order is achieved locally due to an accumulation of particles in a band already before the homogeneous solution becomes polarly ordered.
We expect, that the system size dependence of the transition noise becomes less pronounced for higher densities, because in this case, the density fluctuations are less important.

In order to confirm the applicability of the ring-kinetic theory, we apply it also at different densities.
We formally assumed the limit $N\rightarrow \infty$ in the derivation of the kinetic equations, in particular in the number of neighbor distribution.
For small densities we require large systems in order to obtain reasonably large particle numbers $N$.
However, for those large systems the numerical solution of the ring-kinetic equations becomes computationally to expensive with the Fourier techniques we are using.
Therefore we focus on larger densities $C_1=3$ and $C_1=5$.
For those parameters, the ring-kinetic theory becomes unstable when approaching the flocking transition.
The reason is, that the number of neighbor distribution becomes unphysical, producing negative probabilities.
This artefact is caused by the fact that in reality, higher order correlations have to be taken into account in order to predict the correct number of neighbor distribution.
Ignoring them can lead to negative prefactors $w_k$ in the ring-kinetic equations \eqref{eq:pnormalized} and \eqref{eq:g2_dynamics} that make the equation unstable.
In order to fix this problem we require some information about higher order correlations, such that a reasonable (physical) number of neighbor distribution can be used.
In particular we need the local integrals over the correlation functions:
\begin{align}
	&C_k:= N^k \int_{}^{} G_k(1,...,k)\prod_{l=1}^k \theta(R-|\mathbf{r}_l|)\diff l,
	\label{eq:corrcircle}\\
	&D_k:= N^{k-1}\int G_k(1,2,\dots, k)\diff 1 \prod_{l=2}^k\theta(R-|\mathbf{r}_1-\mathbf{r}_l|)\diff l,
\label{eq:definitionDcoefficients}
\end{align}
see \cite{KSZI20} for details.
We are estimating the next order coefficients $C_3$ and $D_3$ using a closure ansatz presented in the next section.

\section{Closure Relation\label{sec:closure_relation}}

In order to obtain an estimate for the three particle correlation function $G_3$ we employ the closure ansatz
\begin{align}
	P_3(1, 2, 3)= &\Psi(1, 2) \Psi(2, 3) + \Psi(1, 3) \Psi(2, 3)
	\notag
	\\
	&+ \Psi(1, 3) \Psi(1, 2),
	\label{eq:closureansatz1}
\end{align}
where $\Psi(1, 2)$ is a symmetric function
\begin{align}
	\Psi(1, 2) = \Psi(2,1)
	\label{eq:symmetry_psi}
\end{align}
that satisfies
\begin{align}
	\int_{}^{} \diff 2 \Psi(1, 2) = \Gamma =const.
	\label{eq:psi_gamma}
\end{align}
We assume furthermore that $\Psi(1,2)$ is translational invariant. 
Similar closures have been used in astronomy or plasma physics, see e.g. \cite{DP77, White79, NR65}.
Here, the ansatz is motivated by the limit of small noise. There, nearby particles are strongly aligned. If particle one is aligned with particle two and particle two is aligned with particle three, then particle one will be also aligned with particle three. 
For finite noise however, the quality of the ansatz is not evident a priori. It can be justified only when it leads to reasonable results.

Integrating Eq.~\eqref{eq:closureansatz1} over all degrees of freedom fixes $\Gamma$ as
\begin{align}
	\Gamma= \frac{1}{L\sqrt{6\pi}}.
	\label{eq:closure_gamma}
\end{align}
Integrating Eq.~\eqref{eq:closureansatz1} over the degrees of freedom of particle three leads to
\begin{align}
	\Psi(1, 2) = \frac{L\sqrt{6\pi}}{2} [P_2(1, 2) - \int \diff 3 \Psi(1, 3) \Psi(2, 3)].
	\label{eq:closure_iteration}
\end{align}
For a given $P_2$ we can iterate this equation in order to solve for $\Psi$.
Starting with the initial guess $\Psi_0=const. = \frac{1}{\sqrt{3}2\pi L^2}$, the iterative procedure converges very fast, usually within one or two iterations.
Having solved for $\Psi$, we obtain $P_3$ from the ansatz \eqref{eq:closureansatz1} and $G_3$ according to Eq.~\eqref{eq:g3}.
In principle, we could calculate an additional collision term with $G_3$ in the kinetic equation of $g_2$ \eqref{eq:g2_dynamics}.
Here however, we neglect this term, assuming that the angular dependence of $G_3$ is sufficiently small.
We consider only the effect of $G_3$ on the number of neighbor distribution.
That means we calculate $C_3$ and $D_3$ according to Eqs. \eqref{eq:corrcircle} and \eqref{eq:definitionDcoefficients} and use the number of neighbor distribution of \cite{KSZI20} to calculate the prefactors in the kinetic equations \eqref{eq:pnormalized} and \eqref{eq:g2_dynamics}.

\begin{figure}[h]
	\begin{center}
		\includegraphics{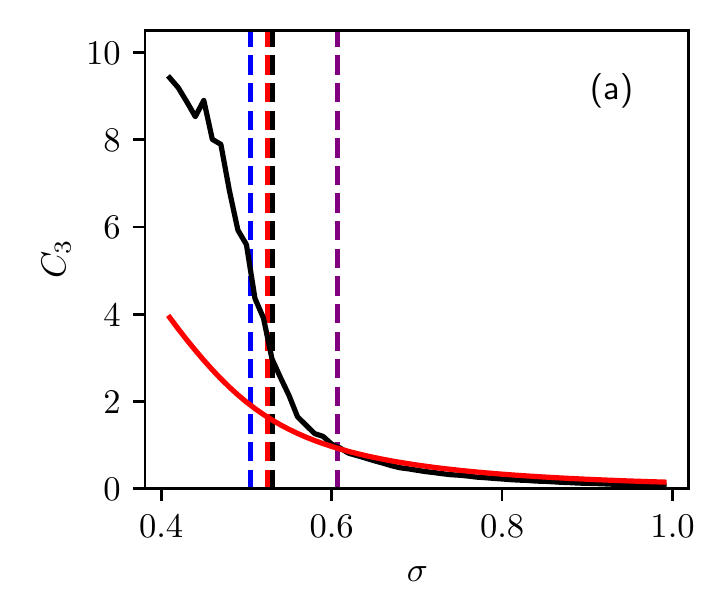}\\
		\includegraphics{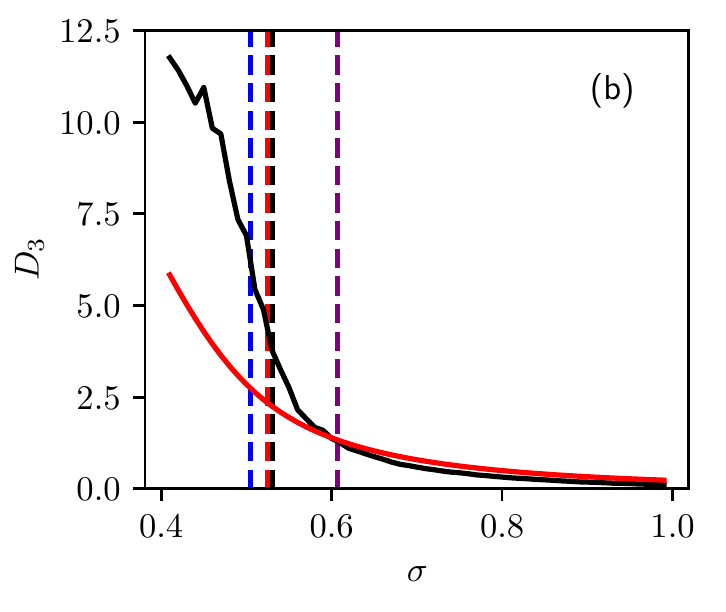}
	\end{center}
	\caption{Local spatial three particle correlation parameters $C_3$ (a) and $D_3$ (b) as defined in Eq.~\protect\eqref{eq:correlation_coefficients} compared for direct simulations (black line) and a kinetic theory including a $g_3$-closure (red line).
	The dashed vertical lines shows the onset of flocking measured in direct simulations (black), in the ring-kinetic theory (blue), ring-kinetic theory with closure (red) and in mean field theory (purple).
	System parameters are as in Fig.\ref{fig:c2_d2}.}
	\label{fig:c3_d3}
\end{figure}

Considering the parameters of section \ref{sec:numerics}, with density $C_1=1$ we compare the measured values of the three particle correlation parameters $C_3$ and $D_3$ with the results of the kinetic theory in Fig.~\ref{fig:c3_d3}.
For large noise, the relative difference between theory and simulation is large.
However, in that case the influence of three particle correlations is negligible any way.
For smaller noise, coming closer to the flocking transition, the closure becomes better and agrees very well with the direct simulations.
Even closer to the transition, the closure clearly underestimates the correlations.
However, the agreement is satisfactory almost up to the transition.
The introduction of the closure relations also improves the agreement of pair correlations with the simulation, cf. Fig.~\ref{fig:c2_d2}.
Furthermore, the prediction of the transition noise strength is further improved to $\sigma_{c, \text{closure}}=0.525(5)$.
It agrees with the value from direct simulations within the reached uncertainty.

\subsection{Density Dependence}

\begin{figure}[h]
	\begin{center}
		\includegraphics[width=0.45\columnwidth]{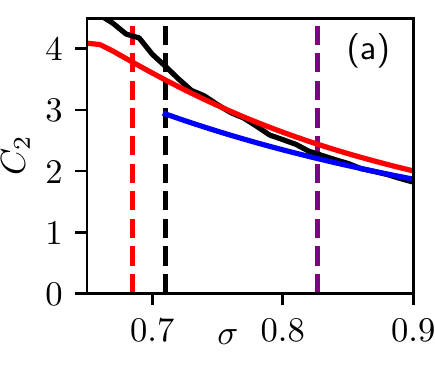}
		\includegraphics[width=0.45\columnwidth]{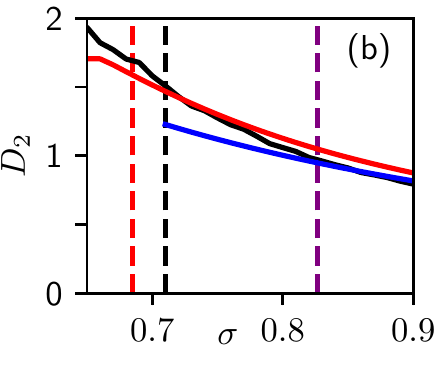}\\
		\includegraphics[width=0.45\columnwidth]{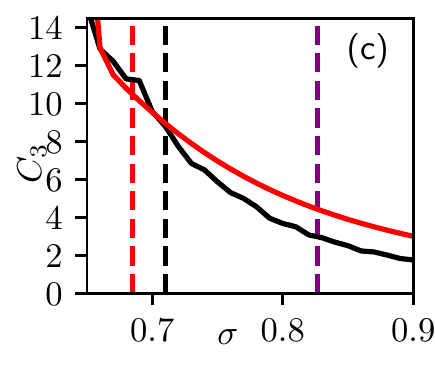}
		\includegraphics[width=0.45\columnwidth]{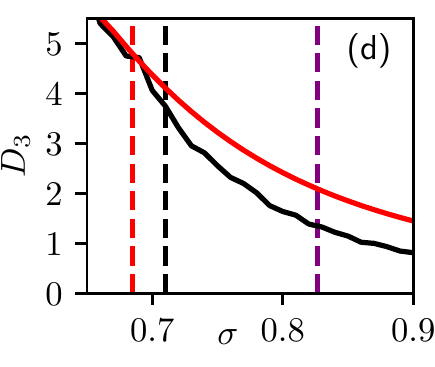}\\
	\end{center}
	\caption{Spatial pair correlation parameters $C_2$ (a) and $D_2$ (b) and three particle correlation parameters $C_3$ (c) and $D_3$ (d) measured in agent-based simulations (black line), within pure ring-kinetic theory (blue line) and within kinetic theory including a three particle closure (red line). 
	The vertical dashed lines display the onset of flocking in kinetic theory with closure at $\sigma_{c, closure}=0.685$ (red), agent-based simulations at $\sigma_{c, sim}=0.71$ (black) and mean field theory at $\sigma_{c, mf}=0.827$ (purple). Sytem parameters are $C_1=\rho \pi R^2=3$, $v=R=1$, $N=300$.}
	\label{fig:M3_c2_d2_c3_d3}
\end{figure}
\begin{figure}[h]
	\begin{center}
		\includegraphics[width=0.45\columnwidth]{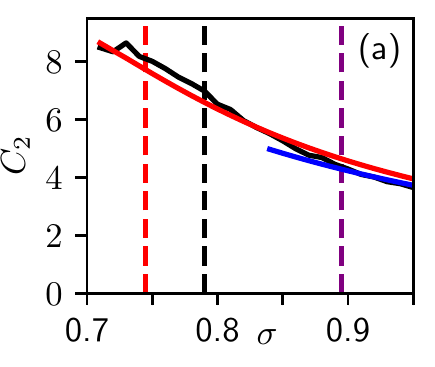}
		\includegraphics[width=0.45\columnwidth]{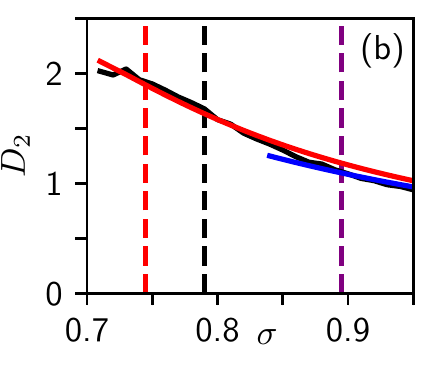}\\
		\includegraphics[width=0.45\columnwidth]{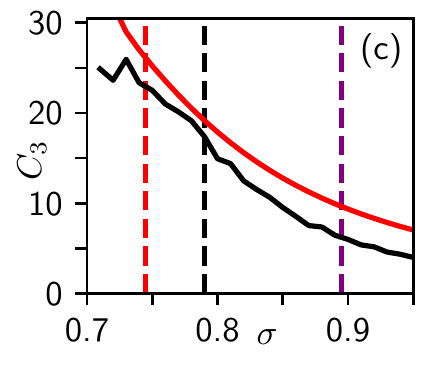}
		\includegraphics[width=0.45\columnwidth]{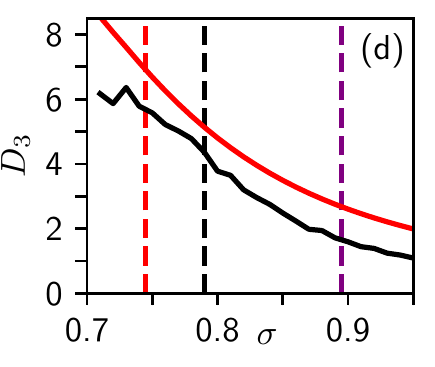}\\
	\end{center}
	\caption{Spatial pair correlation parameters $C_2$ (a) and $D_2$ (b) and three particle correlation parameters $C_3$ (c) and $D_3$ (d) measured in agent-based simulations (black line), within pure ring-kinetic theory (blue line) and within kinetic theory including a three particle closure (red line).
	The vertical dashed lines display the onset of flocking in kinetic theory with closure at $\sigma_{c, closure}=0.745$ (red), agent-based simulations at $\sigma_{c, sim}=0.79$ (black) and mean field theory at $\sigma_{c, mf}=0.895$ (purple). Sytem parameters are $C_1=\rho \pi R^2=5$, $v=R=1$, $N=500$.}
	\label{fig:M5_c2_d2_c3_d3}
\end{figure}
For larger densities $C_1=3$ and $C_1=5$ we find similar results.
The pair correlations predicted by the kinetic theory agree within about 10 \% with the direct simulations up to the flocking transition, cf. Fig.~\ref{fig:M3_c2_d2_c3_d3}a and b and Fig.~\ref{fig:M5_c2_d2_c3_d3}a and b.
Also the three particle correlations predicted by the closure ansatz have the correct order of magnitude, see Fig.~\ref{fig:M3_c2_d2_c3_d3}c and d and Fig.~\ref{fig:M5_c2_d2_c3_d3}c and d.

For $C_1=3$ we find the transition noise strength at $\sigma_{c, \text{closure}}=0.685(5)$ compared to the value measured in direct simulations $\sigma_{c, \text{sim}}=0.71(1)$ and the mean field value $\sigma_{c, \text{mf}}=0.827$. 
For $C_1=5$ the obtained transition noise strengths are $\sigma_{c, \text{closure}}=0.745(5)$, $\sigma_{c, \text{sim}}=0.79(1)$ and $\sigma_{c, \text{mf}}=0.895$.
In summary, we find that the role of three particle correlations increases for larger densities at the flocking transition.
Therefore one expects that also higher order correlations become important.
This explains that deviations between theory and agent-based simulations increase for large densities.
Nevertheless, we still find satisfactory quantitative agreement with the simulations up to $C_1=5$.
The introduction of a closure relation for spatial three particle correlations enlarged the parameter region of applicability of the kinetic theory significantly compared to the pure ring-kinetic theory without closure.

\subsection{Velocity Dependence}

\begin{figure}[h]
	\begin{center}
		\includegraphics[width=0.45\columnwidth]{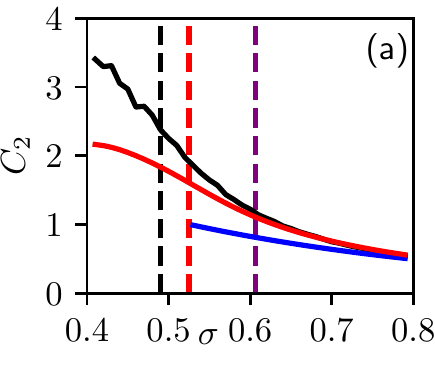}
		\includegraphics[width=0.45\columnwidth]{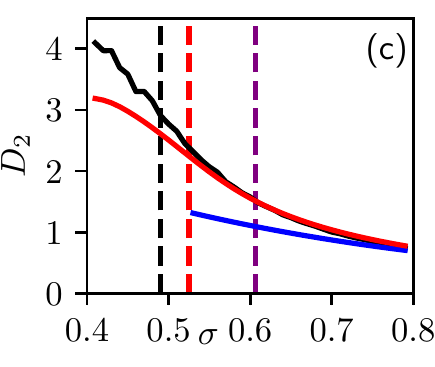}\\
		\includegraphics[width=0.45\columnwidth]{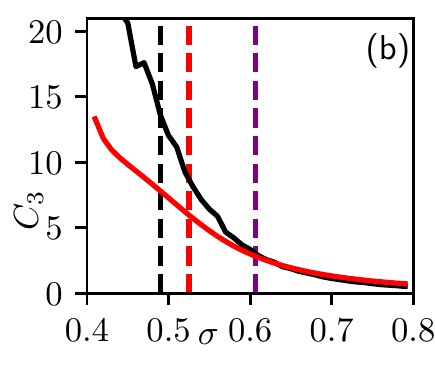}
		\includegraphics[width=0.45\columnwidth]{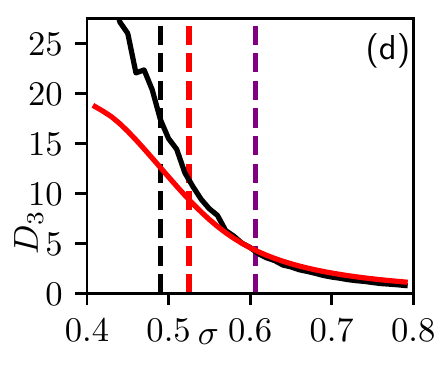}\\
	\end{center}
	\caption{Spatial pair correlation parameters $C_2$ (a) and $D_2$ (b) and three particle correlation parameters $C_3$ (c) and $D_3$ (d) measured in agent-based simulations (black line), within pure ring-kinetic theory (blue line) and within kinetic theory including a three particle closure (red line).
	The vertical dashed lines display the onset of flocking in kinetic theory with closure at $\sigma_{c, closure}=0.525$ (red), agent-based simulations at $\sigma_{c, sim}=0.49$ (black) and mean field theory at $\sigma_{c, mf}=0.607$ (purple). Sytem parameters are $C_1=\rho \pi R^2=1$, $v=0.5$, $R=1$, $N=509$.}
	\label{fig:v0_5_c2_d2_c3_d3}
\end{figure}

\begin{figure}[h]
	\begin{center}
		\includegraphics[width=0.45\columnwidth]{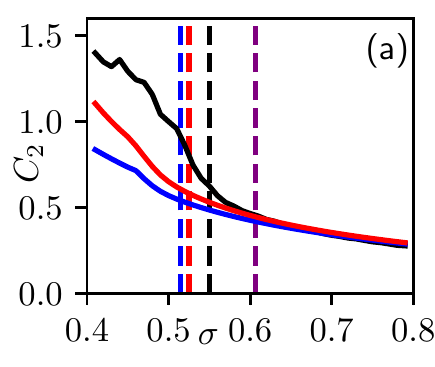}
		\includegraphics[width=0.45\columnwidth]{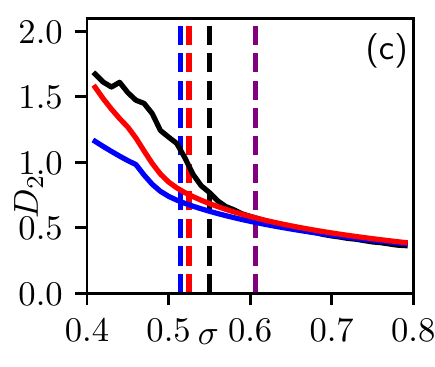}\\
		\includegraphics[width=0.45\columnwidth]{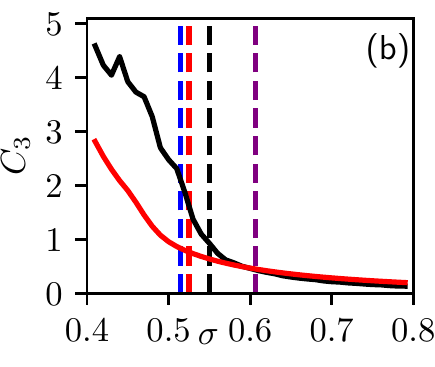}
		\includegraphics[width=0.45\columnwidth]{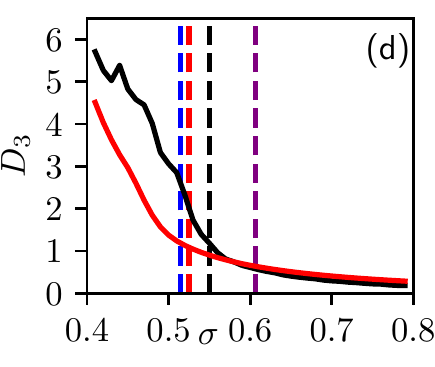}\\
	\end{center}
	\caption{Spatial pair correlation parameters $C_2$ (a) and $D_2$ (b) and three particle correlation parameters $C_3$ (c) and $D_3$ (d) measured in agent-based simulations (black line), within pure ring-kinetic theory (blue line) and within kinetic theory including a three particle closure (red line). 
	The vertical dashed lines display the onset of flocking in pure ring-kinetic theory at $\sigma_{c, rk}=0.515$ (blue), kinetic theory with closure at $\sigma_{c, closure}=0.525$ (red), agent-based simulations at $\sigma_{c, sim}=0.55$ (black) and mean field theory at $\sigma_{c, mf}=0.607$ (purple). System parameters are $C_1=\rho \pi R^2=1$, $v=1.5$, $R=1$, $N=509$.}
	\label{fig:v1_5_c2_d2_c3_d3}
\end{figure}

We study the parameter set of Fig.~\ref{fig:c2_d2} at $C_1=\rho \pi R^2=1$ for different velocities.
For a smaller velocity of $v=0.5$ we find larger correlations, see Fig.\ref{fig:v0_5_c2_d2_c3_d3}.
This is expected because particles have more time to interact and align at smaller velocities.
The predictions of the pure ring-kinetic theory do not agree as well as for $v=1$ similar as for larger densities.
These deviations are very likely caused by higher order correlations that need to be taken into account.
As for larger densities, the ring-kinetic theory is unstable for small noise strengths and the description breaks down already above the flocking transition.
Including the closure improves the kinetic theory significantly.
However, there are significant deviations of pair- and three particle correlations when the flocking transition is approached.
Nevertheless, the flocking transition predicted by our kinetic theory is a clear improvement compared to mean field theory.

For a larger velocity, $v=1.5$ we find smaller spatial correlations as expected, see Fig.~\ref{fig:v1_5_c2_d2_c3_d3}.
The spatial pair correlations agree also very well between ring-kinetic theory and agent-based simulations almost until the onset of flocking.
Here, the ring-kinetic theory is stable and predicts the flocking the transition at slightly too small noise.
Thus, the closure improves the results slightly.
In principle, we would expect better agreement due to the smaller correlations.
However, only the pure spatial correlations decrease for larger velocities,
but the angular correlations increase because particles can only stay close to each other for a long time if they are aligned at high velocities.
For even larger velocity (e.g. $v=2$) we observe very strong angular correlations that suppress the onset of flocking. That means we find no onset of flocking within the ring-kinetic theory (with or without closure) although the spatial correlations are predicted reasonably well. 
We believe that higher order angular correlations need to be taken into account for high velocities.
The next step in this direction will be the incorporation of $g_3$-collision integrals into Eq.~\eqref{eq:g2_dynamics}, where $g_3$ is still determined by the closure \eqref{eq:closureansatz1}.

\section{Discussion\label{sec:discussion}}

We consider polarly aligning self-propelled point particles in two dimensions.
The investigated models are in the spirit of the famous Vicsek models.
However, they follow a continuous time dynamics given by a system of generalized Langevin equations.
For technical reasons it would be desirable to introduce only pair interactions.
But in that case the model behaves qualitatively different from the Vicsek model: in the ordered phase, no bands or cross sea patterns are observed. 
Instead, strongly aligned high density clusters are formed.

In order to observe a behavior qualitatively equivalent to the Vicsek model, we need to introduce $N$-particle-, that is non-additive interactions as pointed out recently in \cite{CSP21, Stroteich19}.

The presence of $N$-particle interactions seriously complicates matters.
As a consequence, the kinetic equations contain weight factors that are expectation values of the number of neighbor distribution.
It was discovered recently, how this distribution can be calculated exactly even in the presence of many particle correlations \cite{KSZI20}.
Employing this predecessor work, we were able to handle the complications arising from the non-additive interactions.
From a technical point of view, the incorporation of $N$-particle interactions into ring-kinetic theories is a significant development that can be of interest also for other (possibly passive) systems with interactions of this type.

Truncating the BBGKY-hierarchy after the second equation we obtain the time evolution equations for the one particle distribution and the pair correlation function, neglecting higher order correlations.
We solve those equations numerically, transforming both, spatial and angular dependence, to Fourier space.
We compare the resulting steady states with direct, agent-based simulations of the Langevin equations.
In the disordered phase and not too close to the onset of flocking, we find excellent agreement between ring-kinetic theory and direct simulations.

Reducing the noise strength, we find the onset of flocking in the ring-kinetic theory. 
Compared to the homogeneous mean field theory, the transition is shifted towards smaller noise.
This shift is caused by positive angular pair correlations that stabilize the disordered phase.
The same effect is seen in direct simulations.
For $C_1=\rho \pi R^2=1$ we find good quantitative agreement of the onset of flocking between ring-kinetic theory and agent-based simulations.
For larger densities, close to the onset of flocking, the ring-kinetic equations become unstable because the weights depending on the number of neighbor distribution are predicted wrong by the ring-kinetic theory.
This is not too surprising as it is known that higher order correlations are required in order to predict the correct number of neighbor distribution and thus the correct weights within the ring-kinetic equations \cite{KSZI20}.

In order to enlarge the applicability of the kinetic theory we introduce a closure ansatz.
With that relation we calculate the three particle correlation function $g_3$ from the single particle distribution and the pair correlations.
In this work, we do not consider the full effect of the three particle correlations, but only its influence on the number of neighbor distribution and hence on the weights in the kinetic equations.
Phrased differently, we consider only the effect of three particle correlations on the spatial distribution and we neglect angular three particle correlations that are entering the kinetic equations by additional collision integrals.

Far in the disordered phase, for large noise, the closure ansatz does not agree very well with direct simulations.
However, in that region three particle correlations can be ignored as discussed above.
At about the onset of flocking, the closure ansatz for the spatial three particle correlations agrees within about $20\%$ with direct simulations.
That means it is a considerable improvement compared to the neglect of those correlations.
Furthermore, the introduction of the closure significantly enlarges the parameter regime where spatial pair correlations agree quantitatively with direct simulations.

Employing the closure ansatz we are able to describe the onset of flocking within kinetic theory also for larger densities such as $C_1=\rho \pi R^2=3, 5$ with deviations of the flocking noise strength of about $4\%$ and $6\%$.
In comparison, the deviations of the transition noise in mean field theory are $16\%$ and $13\%$, respectively.

We also consider the dependence on the particle velocity and find good agreement between ring-kinetic theory and simulations at large noise for all considered velocities.
Close to the onset of flocking, small velocities increase in particular spatial correlation whereas high velocities suppress spatial correlations but favor angular correlations.
This is intuitive because at high speed, particles remain close to each other for long times only if they have roughly the same velocity, and for small velocities, particles that interact have a long time to align.
Both effects lead to deviations of the predicted onset of flocking in the kinetic theory with three particle closure.
However, the results of the kinetic theory are still a major improvement over mean field theory.
There is further potential to improve the theory for high velocities by incorporating $g_3$ collision integrals in the $g_2$ equation by means of the presented closure ansatz. 

In general, it can be seen that the ring-kinetic theory (possibly extended by a closure relation estimating $g_3$) gives quantitatively good results as long as higher order correlations are not too large.
Far in the disordered phase (for large noise), this is always the case.
Close to the flocking transition we observe larger spatial correlations for higher particle densities or small velocities and large angular correlations for high velocities.

Note, in this work we considered relatively small systems consisting of only a few hundred particles.
The reason is that the numerical solution of the kinetic equations is too time consuming for much larger systems because the number of necessary spatial Fourier modes is proportional to the area of the system.
Our analysis predicts that the correlations are mainly localized within the interaction region.
That suggests the use of different spatial decompositions like e.g. into Hermite functions instead of Fourier modes.

\begin{acknowledgments}
The authors gratefully acknowledge the GWK support for funding this project by providing computing time through the Center for Information Services and HPC (ZIH) at TU Dresden on the HRSK-II. The authors gratefully acknowledge the Universit\"atsrechenzentrum Greifswald for providing computing. We thank Aurelio Patelli, Fernando Peruani and Sven Stroteich for valuable discussions.
\end{acknowledgments}

\begin{appendix}

\section{Fourier Transform\label{sec:fourier}}
\subsection{Time Evolution Equations}

We solve the time evolution equations \eqref{eq:pnormalized} and \eqref{eq:p2} numerically in Fourier space, transforming both, spatial and angular coordinates.
This has the huge advantage that all appearing integrals (also high dimensional ones) can be solved analytically.
We employ the following Fourier ansatz for the one particle distribution and the pair correlation function. 
\begin{align}
	p(\phi)= \sum_{k} A_{k} \exp(ik\phi),
	\label{eq:fourieransatzp1}
\end{align}
\begin{align}
	g_2(\phi_1, \phi_2, \Delta)= &\sum_{k, l, m, n} F_{k, l, m, n} \exp(ik\phi_1) \exp(il\phi_2)
	\notag
	\\
	&\times \exp(im\Delta_x 2\pi/L) \exp(in\Delta_y 2\pi/L).
	\label{eq:fourieransatzg1}
\end{align}
The number of neighbor distribution that determines the weights in Eqs.~\eqref{eq:pnormalized} and \eqref{eq:p2} is known analytically up to a discrete Fourier transform \cite{KSZI20}.
It depends only on the integrals \eqref{eq:correlation_coefficients} when three particle and higher order correlations are neglected.
Those integrals can be evaluated in Fourier space.
It is useful to introduce the abbreviations
\begin{align}
	K_{m, n}:=&\frac{1}{L^2} \int_{-L/2}^{L/2}\int_{-L/2}^{L/2}e^{im\frac{2\pi}{L}x}e^{in\frac{2\pi}{L}y}
	\notag
	\\
	&\times\theta(R-|\mathbf{x}|)\diff x \diff y
	\notag
	\\
	=& \frac{R}{L \sqrt{m^2 + n^2}}
	\notag
	\\
	&\times J_{1}\Big(\frac{2\pi}{L}R \sqrt{m^2+n^2}\Big),
	\label{eq:fourierhat}
\end{align}
\begin{align}
	Q_{klmn}^{\pm} := \sum_{s,t} F_{klst}K_{m\pm s,n\pm t} 
	\label{eq:defQ}
\end{align}
and
\begin{align}
	R_{klmn}^{\pm} := \sum_{s,t}F_{klst}K_{m\pm s,n\pm t}C_{s,t},
	\label{eq:defR}
\end{align}
where $J_1$ is the Bessel function of the first kind.
With those abbreviations one obtains
\begin{align}
	C_2 =& 4 \pi^2 N^2 R^+_{0000},
	\\
	D_2=& 4\pi^2 N Q^+_{0000}.
	\label{eq:correlation_coefficients_from_fourier}
\end{align}
Thus, with the results of \cite{KSZI20}, the weights in Eqs.~\eqref{eq:pnormalized} and \eqref{eq:p2} can be expressed explicitly in terms of the Fourier modes $F_{klmn}$.
Fourier transforming those time evolution equations results in
\begin{align}
	&\partial_t A_k= N w_2 k \pi (Q^+_{k-1,1,0,0}-Q^+_{k+1,-1,0,0})
	\notag
	\\
	&+C_1 w_2 k \pi (A_{k-1}A_{1}-A_{k+1}A_{-1})
	\notag
	\\
	&+N^2 2 \pi^2 k(w_3-w_2)(A_{k-1}R^+_{1,0,0,0}-A_{k+1}R^+_{-1,0,0,0})
	\notag
	\\
	&+C_1 N 2 \pi^2 k (w_3-w_2)(A_1 Q^+_{k-1,0,0,0}-A_{-1}Q^+_{k+1,0,0,0})
	\notag
	\\
	&+N^3 4 \pi^3 k (w_4-2w_3+w_2)
	\notag
	\\
	&\times (Q^+_{k-1,0,0,0}R^+_{1,0,0,0}-Q^+_{k+1,0,0,0}R^+_{-1,0,0,0})
	\notag
	\\
	&- \frac{\sigma^2}{2}k^2 A_k.
	\label{eq:time_evolution_angular_modes}
\end{align}
and
\begin{align}
	\partial_t F_{klmn}= \sum_{i=1}^{36} \kreis{i} + \{k\leftrightarrow l, m\leftrightarrow-m, n\leftrightarrow-n\},
	\label{eq:time_evolution_correlation_modes}
\end{align}
where the terms $\kreis{i}$ are given by
\begin{align}
	\kreis{1}= w_2 \frac{k}{2} (Q^-_{k-1, l+1, m, n} - Q^-_{k+1, l-1, m, n}),
	\label{eq:term1}
\end{align}
\begin{align}
	\kreis{2}= w_2 \frac{k}{2} K_{mn}(A_{k-1}A_{l+1}- A_{k+1}A_{l-1}),
	\label{eq:term2}
\end{align}
\begin{align}
	\kreis{3}= N k w_3 \pi (A_{k-1}R^+_{l, 1, m, n}- A_{k+1}R^+_{l, -1, m, n}),
	\label{eq:term3}
\end{align}
\begin{align}
	\kreis{4}=C_1 \pi k w_3(Q^-_{k-1, l, m, n}A_1 - Q^-_{k+1, l, m, n}A_{-1}),
	\label{eq:term4}
\end{align}
\begin{align}
	\kreis{5}=N\pi k (w_3-w_2)K_{mn} A_l(Q^+_{k-1, 1, 0,0}-Q^+_{k+1,-1, 0,0 }),
	\label{eq:term5}
\end{align}
\begin{align}
	&\kreis{6}= N^2 (w_4-w_3)k 2 \pi^2
	\notag
	\\
	&\times (Q^-_{k-1, l, m, n}R^+_{1, 0, 0, 0}- Q^-_{k+1, l, m, n} R^+_{-1, 0, 0, 0}),
	\label{eq:term6}
\end{align}
\begin{align}
	&\kreis{7}=N^2(w_4-w_3)2 \pi^2 k
	\notag
	\\
	&\times (R^+_{l, 1, m, n} Q^+_{k-1, 0, 0, 0} - R^+_{l, -1, m, n}Q^+_{k+1, 0, 0, 0}),
	\label{eq:term7}
\end{align}
\begin{align}
	&\kreis{8}= N^2 (w_4-w_3)k 2 \pi^2
	\notag
	\\
	&\times R^+_{l, 0, m, n} (Q^+_{k-1, 1, 0, 0}-Q^+_{k+1,-1,0, 0}),
	\label{eq:term8}
\end{align}
\begin{align}
	&\kreis{9}= C_1 N (w_4-w_3)k 2 \pi^2 
	\notag
	\\
	&\times R^+_{l,0,m,n}(A_{k-1}A_{1}- A_{k+1}A_{-1}),
	\label{eq:term9}
\end{align}
\begin{align}
	&\kreis{10}=N^2(w_4-2w_3+w_2)k 2 \pi^2 A_l K_{mn}
	\notag
	\\
	&\times(A_{k-1} R^+_{1,0,0,0}- A_{k+1}R^+_{-1,0,0,0}),
	\label{eq:term10}
\end{align}
\begin{align}
	&\kreis{11}= C_1 N k 2 \pi^2 (w_4-2w_3+w_2)A_l K_{mn}
	\notag
	\\
	& \times (Q^+_{k-1,0,0,0}A_{1} - Q^+_{k+1, 0, 0, 0} A_{-1}),
	\label{eq:term11}
\end{align}
\begin{align}
	&\kreis{12}= N^2 C_1 4 \pi^3 k (w_5-2w_4+w_3)R^+_{l,0,m,n}
	\notag
	\\
	&\times (Q^+_{k-1,0,0,0} A_{1} - Q^+_{k+1, 0,0 ,0} A_{-1}),
	\label{eq:term12}
\end{align}
\begin{align}
	&\kreis{13}= N^3 k 4 \pi^3(w_5-2w_4+w_3) R^+_{l,0,m,n}
	\notag
	\\
	&\times (A_{k-1}R^+_{1,0,0,0}-A_{k+1}R^+_{-1,0,0,0}),
	\label{eq:term13}
\end{align}
\begin{align}
	&\kreis{14}=N^3 k 4 \pi^3(w_5-3w_4+3w_3 -w2) K_{mn} A_{l}
	\notag
	\\
	&\times ( Q^+_{k-1, 0, 0, 0} R^+_{1, 0, 0,0} - Q^+_{k+1, 0, 0, 0} R^+_{-1, 0, 0,0},
	\label{eq:term14}
\end{align}
\begin{align}
	&\kreis{15} = N^4 8 \pi^4 k (w_6 -3 w_5+ 3 w_4 -w_3) R^+_{l, 0, m, n}
	\notag
	\\
	&\times (Q^+_{k-1,0,0,0} R^+_{1,0,0,0} - Q^+_{k+1,0,0,0} R^+_{-1,0,0,0}),
	\label{eq:term15}
\end{align}
\begin{align}
	&\kreis{16}=N k \pi w_2 K_{mn} 
	\notag
	\\
	& \times(A_{k-1}F_{l, 1, -m, -n} - A_{k+1} F_{l,-1,-m,-n}),
	\label{eq:term16}
\end{align}
\begin{align}
	&\kreis{17}= -Nk\pi w_2(A_{k-1} R^+_{l, 1, m, n} - A_{k+1} R^+_{l, -1, m, n})
	\label{eq:term17}
\end{align}
\begin{align}
	&\kreis{18}= C_1 w_2 k \pi(F_{k-1, l, m, n} A_{1} - F_{k+1, l, m, n} A_{-1}),
	\label{eq:term18}
\end{align}
\begin{align}
	&\kreis{19} = - C_1 w_2 k \pi (Q^-_{k-1, l, m, n}A_{1} - Q^-_{k+1, l, m, n} A_{-1}),
	\label{eq:term19}
\end{align}
\begin{align}
	&\kreis{20} = N^2 2 \pi^2 k (w_3-w_2)
	\notag
	\\
	&\times(R^+_{1,0,0,0} F_{k-1, l, m, n}- R^+_{-1,0,0,0} F_{k+1, l, m, n}),
	\label{eq:term20}
\end{align}
\begin{align}
	&\kreis{21}= - N^2 2 \pi^2 k (w_3-w_2)
	\notag
	\\
	&\times(Q^-_{k-1, l, m, n}R^+_{1,0,0,0}- Q^-_{k+1, l, m, n} R^+_{-1,0,0,0}),
	\label{eq:term21}
\end{align}
\begin{align}
	&\kreis{22}= N^2 k 2 \pi^2 (w_3-w_2)K_{mn}
	\notag
	\\
	&\times(Q^+_{k-1, 0, 0, 0}F_{l, 1, -m, -n}- Q^+_{k+1, 0, 0, 0} F_{l, -1, -m, -n}),
	\label{eq:term22}
\end{align}
\begin{align}
	&\kreis{23}= - N^2 k 2 \pi^2 (w_3-w_2)
	\notag
	\\
	&\times (Q^+_{k-1, 0, 0, 0}R^+_{l, 1, m, n}- Q^+_{k+1, 0, 0, 0}R^+_{l, -1, m, n}),
	\label{eq:term23}
\end{align}
\begin{align}
	&\kreis{24} = N^2 k 2 \pi^2 (w_3-w_2) K_{mn}
	\notag
	\\
	&\times F_{l, 0, -m, -n} (Q^+_{k-1, 1, 0, 0} - Q^+_{k+1, -1, 0, 0}),
	\label{eq:term24}
\end{align}
\begin{align}
	&\kreis{25} = - N^2 k 2 \pi^2 (w_3-w_2) R^+_{l, 0, m, n}
	\notag
	\\
	&\times (Q^+_{k-1, 1, 0, 0} - Q^+_{k+1, -1, 0, 0}),
	\label{eq:term25}
\end{align}
\begin{align}
	&\kreis{26} = C_1 N k 2 \pi^2 (w_3-w_2) K_{mn}
	\notag
	\\
	&\times F_{l, 0, -m, -n} (A_{k-1}A_{1}- A_{k+1} A_{-1}),
	\label{eq:term26}
\end{align}
\begin{align}
	&\kreis{27}= - C_1 N k 2 \pi^2 (w_3-w_2) 
	\notag
	\\
	&\times R^+_{l,0,m,n} (A_{k-1}A_{1} - A_{k+1} A_{-1}),
	\label{eq:term27}
\end{align}
\begin{align}
	&\kreis{28} = C_1 N^2 k 4 \pi^3 (w_4-2w_3+w_2)K_{mn}
	\notag
	\\
	&\times F_{l, 0, -m, -n} (Q^+_{k-1, 0, 0, 0} A_{1} - Q^+_{k+1, 0, 0, 0} A_{-1}),
	\label{eq:term28}
\end{align}
\begin{align}
	&\kreis{29} - C_1 N^2 k 4 \pi^3 (w_4-2w_3+w_2)
	\notag
	\\
	&\times R^+_{l, 0, m, n}(Q^+_{k-1, 0, 0, 0}A_{1} - Q^+_{k+1, 0, 0, 0} A_{-1}),
	\label{eq:term29}
\end{align}
\begin{align}
	&\kreis{30} = N^3 k 4 \pi^3 (w_2-2w_3+w_2) K_{mn} 
	\notag
	\\
	&\times F_{l, 0, -m, -n} (A_{k-1} R^+_{1,0,0,0} - A_{k+1} R^+_{-1, 0,0,0}),
	\label{eq:term30}
\end{align}
\begin{align}
	&\kreis{31} = -N^3 k 4 \pi^3 (w_2-2w_3+w_2)  
	\notag
	\\
	&\times R^+_{l, 0, m, n} (A_{k-1} R^+_{1,0,0,0} - A_{k+1} R^+_{-1, 0,0,0}),
	\label{eq:term31}
\end{align}
\begin{align}
	&\kreis{32} = N^4 k 8 \pi^4 (w_5-3w_4 +3w_3 - w_2) F_{l,0,-m,-n}
	\notag
	\\
	&\times K_{mn} (Q^+_{k-1, 0, 0, 0} R^+_{1, 0, 0, 0} - Q^+_{k+1, 0, 0, 0} R^+_{-1, 0, 0, 0}),
	\label{eq:term32}
\end{align}
\begin{align}
	&\kreis{33} = -N^4 k 8 \pi^4 (w_5-3w_4 +3w_3 - w_2) R^+_{l,0,m,n}
	\notag
	\\
	&\times (Q^+_{k-1, 0, 0, 0} R^+_{1, 0, 0, 0} - Q^+_{k+1, 0, 0, 0} R^+_{-1, 0, 0, 0}),
	\label{eq:term33}
\end{align}
\begin{align}
	&\kreis{34} = v m \frac{i \pi}{L} (F_{k-1, l, m, n} + F_{k+1, l, m, n}),
	\label{eq:term34}
\end{align}
\begin{align}
	&\kreis{35} = v n \frac{ \pi}{L} (F_{k-1, l, m, n} - F_{k+1, l, m, n}),
	\label{eq:term35}
\end{align}
\begin{align}
	&\kreis{36} = -\frac{\sigma^2}{2} k ^2 F_{klmn}
	\label{eq:term36}
\end{align}
and $\{k\leftrightarrow l, m\leftrightarrow-m, n\leftrightarrow-n\}$ refers to the same terms but with $k$ and $l$, $m$ and $-m$, $n$ and $-n$ interchanged, respectively.

\subsection{Closure Relation}

In order to evaluate the closure ansatz \eqref{eq:closureansatz1} we consider the function $\Psi$ in Fourier space
\begin{align}
	&\Psi(\phi_1, \phi_2, \Delta)= \sum_{klmn} U_{klmn} \exp(ik \phi_1) 
	\notag
	\\
	&\times\exp(i l \phi_2) \exp(im \Delta_x 2\pi/L) \exp(in \Delta_y 2\pi/L).
	\label{eq:fourier_closure}
\end{align}
It is reasonable to consider scaled modes
\begin{align}
	u_{stuv}= L^3 U_{stuv}.
	\label{eq:reduced_u}
\end{align}
From Eqs.~\eqref{eq:psi_gamma} and \eqref{eq:closure_gamma} it follows that
\begin{align}
	u_{s000}= \delta_{s,0} u_{0000}= \delta_{s,0} \frac{L \Gamma}{2\pi}=\delta_{s,0} \frac{1}{2\pi\sqrt{6\pi}}.
	\label{eq:u_s_zero}
\end{align}
Fourier transforming the iteration equation \eqref{eq:closure_iteration} we obtain
\begin{align}
	u_{klmn}= &\frac{1}{2u_{0000}} \bigg[\frac{1}{8\pi^3}\delta_{k,0}\delta_{l,0}\delta_{m,0}\delta_{n,0} 
	\notag
	\\
	&- \sum_{s} u_{ksmn}u_{-slmn} + \frac{1}{2\pi}F_{klmn} \bigg].
	\label{eq:recursion_u}
\end{align}
Note that this equation is compatible with the normalization condition \eqref{eq:u_s_zero}.
In practice, we start with $u_{klmn}=0$ except for $u_{0000}$ that is given by Eq.~\eqref{eq:u_s_zero}.
Then we iterate Eq.~\eqref{eq:recursion_u} to calculate $u_{klmn}$, that is the closure ansatz function $\Psi$.
In the parameter regime we investigated, the recursion converges very fast. 
It almost reaches its fixed point already after two iterations.
We used five iterations after which we reach perfect convergence within our numerical accuracy.

Once $\Psi$ is known, the spatial three particle correlation parameters $C_3$ and $D_3$ can be calculated via
\begin{align}
	&C_3+3C_1C_2+C_1^3=N^3 \int_{}^{} P_3(1,2,3) \diff 1 \diff 2 \diff 3
	\notag
	\\
	&=3 N^3 \int_{}^{} \Psi(1,2) \Psi(2,3) \theta_1 \theta_2 \theta_3 \diff 1 \diff 2 \diff 3
	\notag
	\\
	&= 3 \rho^3 \int_{}^{} L^6 \Psi(1,2) \Psi(2,3) \theta_1 \theta_2 \theta_3 \diff 1 \diff 2 \diff 3
	\notag
	\\
	&= 3 \rho^3 \int_{}^{} \sum_{stuv} \exp(is\phi_1) \exp(it\phi_2) \exp(i u (x_2-x_1)2\pi/L) 
	\notag
	\\
	&\times \exp(i (y_2-y_1)2\pi/L)  \sum_{klmn} \exp(ik \phi_2) \exp(il\phi_3) 
	\notag
	\\
	&\times  \exp(im (x_3-x_2)2\pi/L)
	\notag
	\\
	&\times \exp(in (y_3-y_2)2\pi/L) \theta_1\theta_2\theta_3 \diff 1 \diff 2 \diff 3
	\notag
	\\
	&=3 N^3 (2\pi)^3  \sum_{tuvmn} u_{0,t,u,v} u_{-t, 0,m,n} K_{u,v}K_{u-m,v-n}K_{m,n}
	\label{eq:cthree_closure1}
\end{align}
and similar
\begin{align}
	&D_3 + 2 D_2 C_1 + C_2 + C_1^2 = N^2 \int_{}^{} \theta_{12} \theta_{13} G_3(1, 2, 3) \diff 1 \diff 2 \diff 3
	\notag 
	\\
	&=(2\pi)^3N^2 \sum_{lmnuv}u_{0lmn}u_{-l0uv}K_{m-u,n-v}K_{u,v}
	\notag
	\\
	&+(2\pi)^3N^2 \sum_{lmnuv}u_{0lmn}u_{0-luv}K_{m+u,n+v}K_{u,v}
	\notag
	\\
	&+(2\pi)^3N^2 \sum_{lmnuv}u_{-l0mn}u_{l0uv}K_{m,n}K_{u,v}.
	\label{eq:dthree_closure1}
\end{align}
Note that $C_1:=\frac{N}{L^2}\pi R^2$ and $D_1:=1$ per definition.

In sumary, the numerical time evolution works as follows.
We calculate the pair correlation coefficients $C_2$ and $D_2$ according to Eq.~\eqref{eq:correlation_coefficients_from_fourier}.
Next, we calculate the closure ansatz function with the iteration \eqref{eq:recursion_u} and then the three particle correlation coefficients $C_3$ and $D_3$ according to Eqs.~\eqref{eq:cthree_closure1} and \eqref{eq:dthree_closure1}.
Having calculated the correlation coefficients we compute the number of neighbor distribution as described in detail in \cite{KSZI20}.
With this distribution we obtain the weights $w_k$ according to Eq.~\eqref{eq:neighbor_distr_expectation}.
Given the weights we can eventually time evolve the time evolution equations \eqref{eq:time_evolution_angular_modes} and \eqref{eq:time_evolution_correlation_modes} with a simple Euler scheme.

\subsection{Comparison to Agent-based Simulations}

The correlation functions $g(r)$ and $h(r, \alpha, \Delta \phi)$ considered in Sec.~\ref{sec:numerics} can be sampled in agent-based simulations according to Eqs.~\eqref{eq:def_radial_distribution_function} and \eqref{eq:sampling_h_function}.
On the other hand, these functions can be calculated from the Fourier modes of $g_2$ according to
\begin{align}
	g(r)=1+4\pi^2\sum_{m,n}F_{00mn}J_{0}(\frac{2\pi r}{L}\sqrt{m^2+n^2}),
	\label{eq:fourier_g}
\end{align}
\begin{align}
	h(r, \alpha, \Delta \phi)=& (2\pi)^2 \sum_{klmn} \exp(i l \Delta \phi) \exp[-i(k+l)\alpha] 
	\notag
	\\
	& \times \bigg(\frac{n-im}{\sqrt{n^2+m^2}} \bigg)^{k+l} J_{-(k+l)}\bigg(\frac{2\pi}{L}r \sqrt{m^2+n^2} \bigg).
	\label{eq:fourier_h}
\end{align}
Assuming rotational symmetry, it is also possible to revert Eq.~\eqref{eq:fourier_g} that is to calculate the spatial Fourier modes $F_{00mn}$ from the function $g(r)$ according to
\begin{align}
	F_{00mn}=\frac{1}{L^2 2\pi} \int_{}^{}[g(r)-1]r J_0\bigg(\frac{2 \pi r}{L} \sqrt{m^2+n^2}\bigg) \diff r.
	\label{eq:fourier_g_back}
\end{align}
We use this relation to see the impact of the resolution in the Fourier transform on $g(r)$ by Fourier transforming the measured function according to Eq.~\eqref{eq:fourier_g_back} and transforming it back according to Eq.~\eqref{eq:fourier_g}. In that way we achieved the green dashed line in Fig.~\ref{fig:radial_distribution_function}. The solid black line in this figure is a histogram of agent-based simulations according to Eq.~\eqref{eq:def_radial_distribution_function} and the solid blue line is calculated according to Eq.~\eqref{eq:fourier_g}, where the Fourier modes $F_{klmn}$ are the result of the ring-kinetic theory.

\section{Comparison between full Ring-kinetic Theory and Landau Kinetic Theory\label{sec:landau}}

\begin{figure}[H]
	\begin{center}
		\includegraphics{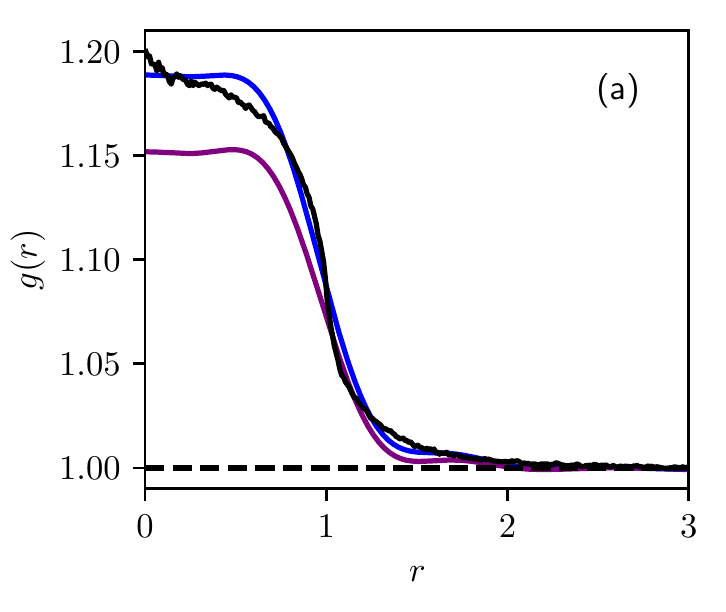}\\
		\includegraphics{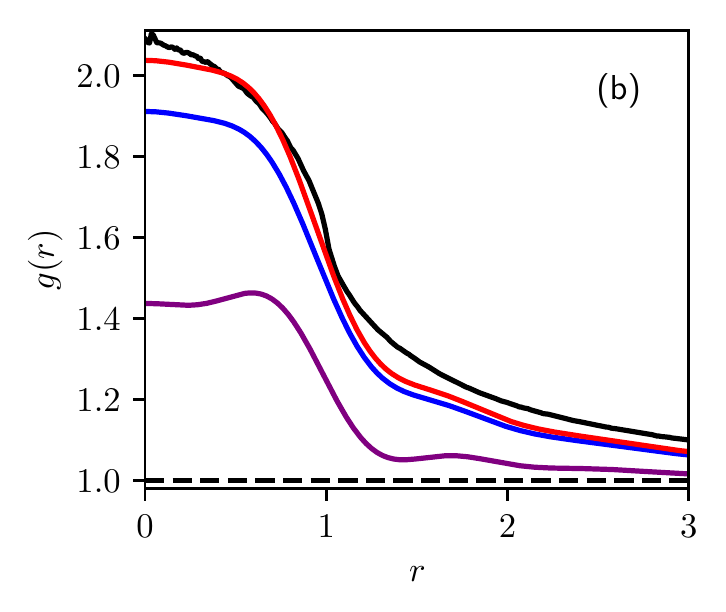}
	\end{center}
	\caption{Radial distribution function $g(r)$ obtained from agent-based simulations (black line), ring-kinetic theory (blue line), kinetic theory with closure (red line) and Landau-kinetic theory according to Eq.~\eqref{eq:patelli2} (purple line). In (a) we used the parameters of Fig.\eqref{fig:radial_distribution_function}, in (b) the noise is smaller $\sigma=0.6$ but still in the disordered phase, other parameters are the same.}
	\label{fig:landau_theory}
\end{figure}

The Landau theory of \cite{Patelli20} considers the model with additive interactions, $w(n)\equiv 1$.
Here, we adopt the theory to the case of non-additive interactions, $w(n)=1/n$ in order to compare it with the present approach.
The equivalent of Eq.~(17) of Ref. \cite{Patelli20} is in our notation
\begin{align}
	&\partial_t g_2(\phi_1, \phi_2, \Delta) =-w_2 \bigg[\tz{\focus \PointC{2} \Arr{\Pa}{\Pc}} + \tz{\focustwo \PointC{1} \Arr{\Pa}{\Pc}}\bigg] 
	\notag
	\\
	&+ v(\cos \phi_1 - \cos \phi_2) \partial_{\Delta_x}g_2(\phi_1, \phi_2, \Delta) 
	\notag
	\\
	&+ v(\sin \phi_1 - \sin \phi_2) \partial_{\Delta_y}g_2(\phi_1, \phi_2, \Delta).
	\label{eq:patelli1}
\end{align}
That is, compared to Eq.~\eqref{eq:g2_dynamics}, all collision integrals on the right hand side that contain $g_2$ are neglected because $g_2$ and the coupling constant are both assumed to be small.
Furthermore, the angular diffusion term $\frac{\sigma^2}{2}(\partial_{\phi_1}^2+\partial_{\phi_2}^2)  g_2(\phi_1, \phi_2, \Delta)$ is neglected because both, $\sigma^2$ and $g_2$, are assumed to be small.
However, we could not make sense out of Eq.~\eqref{eq:patelli1} without the angular diffusion term.
Intuitively, it is clear that the correlations are going to diverge according to Eq.~\eqref{eq:patelli1} because there is a source term but no damping term.
In fact, we find indeed a diverging $g_2$ if we integrate Eq.~\eqref{eq:patelli1} numerically in Fourier space.
Furthermore, we do not see how the principal problem of diverging $g_2$ can be prevented in any parameter limit if the noise term is neglected.
Therefore we did not neglect the angular diffusion and considered instead the time evolution equation
\begin{align}
	&\partial_t g_2(\phi_1, \phi_2, \Delta) =-w_2 \bigg[\tz{\focus \PointC{2} \Arr{\Pa}{\Pc}} + \tz{\focustwo \PointC{1} \Arr{\Pa}{\Pc}}\bigg] 
	\notag
	\\
	&+ v(\cos \phi_1 - \cos \phi_2) \partial_{\Delta_x}g_2(\phi_1, \phi_2, \Delta) 
	\notag
	\\
	&+ v(\sin \phi_1 - \sin \phi_2) \partial_{\Delta_y}g_2(\phi_1, \phi_2, \Delta)
	\notag
	\\
	&+\frac{\sigma^2}{2}(\partial_{\phi_1}^2+\partial_{\phi_2}^2)  g_2(\phi_1, \phi_2, \Delta).
	\label{eq:patelli2}
\end{align}
In Fig.\ref{fig:landau_theory}a we show the pair correlation function $g(r)$ obtained from the Landau kinetic theory (purple line) compared to the results of the full ring-kinetic theory (blue line) and direct agent-based simulations (black line) for the parameters of Fig.\ref{fig:radial_distribution_function} far in the disordered phase at $\sigma=1.5$ (the flocking transition occurs at $\sigma=0.53$).
We find that the qualitative behavior of the pair correlation function is predicted correctly, but it is too small by about $20\%$.
We expect that the results of the Landau kinetic theory improve if much larger noise strengths are considered.
In that case however, the correlations decrease and become less important.

In Fig.\ref{fig:landau_theory}b we display the pair correlation function for the same parameters but smaller noise, $\sigma=0.6$.
The system is still disordered but closer to the flocking transition.
We added also the curve of our kinetic theory including a three particle closure (red line).
Here, we see that the Landau theory is off by more than $50\%$, the ring-kinetic theory is much better but also shows significant deviations and the theory including spatial three particle correlations by a closure ansatz agrees quantitatively very well with the agent-based simulations.
We conclude that the simplified Landau kinetic theory of \cite{Patelli20} is a good approximation far in the disordered regime.
However, it is not suitable to describe the system in the vicinity of the flocking transition.

\end{appendix}

%\bibliography{literatur.bib}

%

\end{document}